\documentclass[aps,prb,twocolumn,superscriptaddress]{revtex4-2}

\usepackage{amsmath,amssymb,amsfonts}
\usepackage{bm, latexsym}
\usepackage[normalem]{ulem}
\usepackage{bbm,times}
\usepackage{braket}
\usepackage{mathtools}
\usepackage{enumerate}

\usepackage{graphicx}
\usepackage{hyperref}
\usepackage{color}

\usepackage{comment}

\newcommand{\cred}[1]{\textcolor{black}{#1}}

\graphicspath{
{./image/}
}

\begin{document}

\makeatletter
\@twocolumntrue
\@firstcolumntrue
\makeatother

\title{Nonequilibrium phase transition of dissipative fermionic superfluids:
Case study of multi-terminal Josephson junctions}
\author{Soma Takemori}
\email{takemori.s.041d@m.isct.ac.jp}
\affiliation{Department of Physics, Institute of Science Tokyo, Meguro, Tokyo 152-8551, Japan}
\author{Kazuki Yamamoto}
\email{kazuki-yamamoto@omu.ac.jp}
\affiliation{Research Institute for Innovation and Co-Creation, Osaka Metropolitan University, Sakai, Osaka 599-8531, Japan}
\affiliation{Department of Physics, Osaka Metropolitan University, Sumiyoshi, Osaka 558-8585, Japan}
\affiliation{Nambu Yoichiro Institute of Theoretical and Experimental Physics (NITEP), Osaka Metropolitan University, Sumiyoshi, Osaka 558-8585, Japan}
\affiliation{Department of Physics, Institute of Science Tokyo, Meguro, Tokyo 152-8551, Japan}

\date{\today}

\begin{abstract}
We investigate nonequilibrium dynamics of a triad of fermionic superfluids connected via Josephson junctions, following sudden switch-on of two-body loss in one of the three superfluids. By formulating the dissipative BCS theory for the Lindblad equation, we find that the superfluid order parameter exhibits a phase rotation, thereby giving rise to three types of dc Josephson currents corresponding to different junctions. 
We demonstrate that, when the tunneling amplitude $V_{31}$ between superfluids without two-body loss is weak, a two-step \cred{nonequilibrium phase transition} characterized by the vanishing dc Josephson currents occurs: dissipation first induces the \cred{nonequilibrium phase transition} by making one dc Josephson current finite, while further increasing dissipation makes this remaining dc Josephson current vanish. By contrast, when $V_{31}$ is strong, dissipation induces the \cred{nonequilibrium phase transition} in which all dc Josephson currents simultaneously vanish. An analytical study based on a simplified model further supports this observation.
\end{abstract}

\maketitle

\section{Introduction}\label{sec_intro}
Collective excitations of fermionic superfluids have attracted significant interest in condensed matter physics~\cite{anderson58,leggett66,volkov73,littlewood82,andreev04,gurarie09,endres12,seibold22}. For example, the amplitude mode of the superfluid order parameter is called the Higgs mode~\cite{sooryakumar80,littlewood81,matsunaga14,krull14,tsuji15,krull16}, the phase mode is known as the Nambu-Goldstone mode~\cite{nambu60,goldstone61,goldstone62},
and the relative phase mode is referred to as the Leggett mode~\cite{leggett66}. A paradigmatic research field for collective excitations is the nonequilibrium quench dynamics of fermionic superfluids, where order parameters exhibit oscillations or characteristic decay~\cite{barankov04,yuzbashyan05,yuzbashyan06,barankov06_sync,barankov06_select,mazza12,yuzbashyan15,hannibal15,mazza17,kettmann17,hannibal18_vanish,hannibal18_pers,ojeda19,seibold20,barresi23,collado23}, and transitions between dynamical phases have been observed depending on the initial conditions and quench parameters~\cite{young24}. Importantly, thanks to the experimental advances in ultracold atoms, nonequilibrium dynamics of the fermionic superfluid has been realized \cite{harrison21, dyke21, cabrera25}, including the observation of the Higgs mode~\cite{behrle18, kell24, dyke24, breyer25}.

On the other hand, dissipative dynamics in open quantum systems has recently attracted a lot of attention~\cite{daley14, ashida20,fazio24,syassen08,yan13,barontini13,zhu14,patil15,labouvie16,luschen17,tomita17,sponselee18,tomita19,corman19,bouganne20,mark20,benary22,honda23,huang23,huang25}.
In systems coupled to environments, dissipation can induce quantum phenomena that are inaccessible in equilibrium settings~\cite{diehl11,ashida16,yamamoto22,yamamoto23local,dai23,stefanini25,yamamoto25measure,gao25fate}, such as nonequilibrium phase transitions~\cite{diehl10insta,honing12,sieberer13,Kessler12, Minganti18,nakagawa21,yamamoto24} and quantum transport in nonequilibrium steady states~\cite{yamamoto20,secli21,visuri22,visuri23non,visuri23dc,muraev24,gievers24,ganguly24,li25}. For example, in a degenerate Fermi gas in one-dimensional junctions subject to one- and two-body losses, anomalous saturation of the thermal and spin conductance has been observed~\cite{huang25}, and dissipative steady states characterized by the unconventional density profile have been investigated~\cite{visuri22}.

Notably, the high controllability of dissipation in ultracold atoms has stimulated the investigations of loss-induced fermionic superfluidity~\cite{yamamoto21,nava23,nava24,gao25dyn,yamamoto19,takemori24honey,takemori25}. For instance, power-law decay of the particle density of fermionic superfluids is studied in the long-time limit using variational methods~\cite{mazza23,filice25}, and BCS theory for a single superfluid Josephson junction shows that dissipation can induce a \cred{nonequilibrium phase transition} in Cooper-pair transports~\cite{yamamoto21}. In particular, such Josephson junctions provide a unique experimental platform for directly probing the relative phase mode of superfluid order parameters and associated Josephson effects reflecting macroscopic quantum coherence in degenerate Fermi gases~\cite{spuntarelli07,valtolina15,husmann15,krinner17,burchianti18,luick20,kwon20,del21,biagioni24,del25}. Interestingly, in the presence of multi-terminal Josephson junctions, the nonequilibrium phenomena of superfluids become more intricate due to multiple coherence among them, e.g., giving rise to unconventional nonlocal Josephson effects~\cite{riwar16}. Thus, the interplay between dissipation~\cite{yamamoto21} and multiple coherence via Josephson junctions~\cite{scherer07, spuntarelli07,valtolina15,burchianti18,luick20,kwon20,del21,biagioni24,del25} is of particular importance as it realizes rich nonequilibrium properties that are accessible with ultracold atoms in open quantum setups.

In this paper, we investigate the loss-quench dynamics of a triad of fermionic superfluids connected via Josephson junctions, following sudden switch-on of two-body loss in one of the three superfluids. We formulate the dissipative BCS theory for the Lindblad equation and derive the time-evolution equation of the system, as well as analytical formulas for the Josephson currents. We show that dissipation induces a phase rotation of the superfluid order parameter, leading to three types of dc Josephson currents corresponding to different junctions. Importantly, when the tunneling amplitude $V_{31}$ between superfluids without two-body loss is weak, we identify a two-step \cred{nonequilibrium phase transition} marked by the vanishing dc Josephson currents: dissipation initially induces the \cred{nonequilibrium phase transition} by rendering one dc current finite, whereas further increase in dissipation leads to the complete disappearance of all dc currents. By contrast, for strong $V_{31}$, the \cred{nonequilibrium phase transition}, characterized by the simultaneous vanishing of all dc Josephson currents, emerges as dissipation increases. The \cred{nonequilibrium phase transition} is further investigated analytically using a calculation of a simplified model.

The rest of this paper is organized as follows. In Sec.~\ref{sec_Model}, we formulate the dissipative BCS theory for dissipative fermionic superfluids connected via multi-terminal Josephson junctions. In Sec.~\ref{sec_DPT}, we present numerical results of the loss-induced nonequilibrium dynamics of fermionic superfluids. Section~\ref{sec_Simp} is devoted to the analysis using a simplified theoretical model. Finally, conclusions are given in Sec.~\ref{sec_Conc}.

\section{Model and methods}\label{sec_Model}
In this section, we formulate the dissipative BCS theory for multi-terminal Josephson junctions and obtain the formula for the Josephson currents by using Anderson's pseudospin representation.
\subsection{Dissipative BCS theory for multi-terminal Josephson junctions}
\begin{figure}[tbh]
    \centering
    \includegraphics[width=8cm]{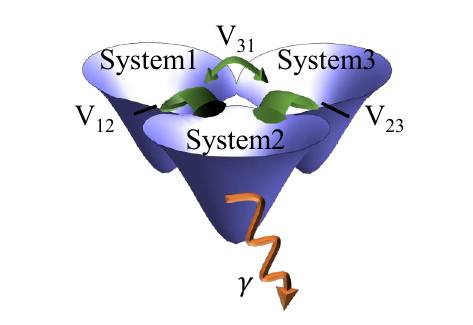}
    \caption{Schematic image of a triad of fermionic superfluids coupled via Josephson junctions. The amplitude of the Cooper-pair tunneling between system $\nu\:(=1,2,3)$ and system $\mu\:(=2,3,1)$ is given by $V_{\nu\mu}$. The two-body loss is introduced in system 2 with rate $\gamma$.}
    \label{fig_schematic}
\end{figure}

We consider a triad of fermionic superfluids connected via Josephson junctions, where two-body loss is introduced in one of the three superfluids (see Fig.~\ref{fig_schematic} for the schematic figure). We consider the pairing Hamiltonian for $s$-wave fermionic superfluids with the Cooper-pair tunneling~\cite{leggett66,krull16,murotani17,yamamoto21} as
\begin{align}
H_\mathrm{sys} = \sum_{\nu}H_{\nu} + H_{\text{tun}},
\end{align}
where
\begin{align}
&H_{\nu} = \sum_{\bm{k}\sigma}\epsilon_{\bm{k}}c_{\nu\bm{k}\sigma}^{\dagger}c_{\nu\bm{k}\sigma} -U_{R}\sum_{\bm{k},\bm{k}'}c_{\nu\bm{k}\uparrow}^{\dagger}c_{\nu-\bm{k}\downarrow}^{\dagger}c_{\nu-\bm{k}'\downarrow}c_{\nu\bm{k}'\uparrow} , \label{tri_SFnuHami_eq}
\end{align}
is the Hamiltonian of system $\nu\:(=1,2,3)$, and
\begin{align}
    H_{\text{tun}} = - \sum_{\langle\nu,\mu\rangle}\sum_{\bm{k}\bm{k}'}&\left[\frac{V_{\nu\mu}}{N_{0}}c_{\nu\bm{k}\uparrow}^{\dagger}c_{\nu-\bm{k}\downarrow}^{\dagger}c_{\mu-\bm{k'}\downarrow}c_{\mu\bm{k'}\uparrow} + \text{H.c.} \right] \label{tri_tunHami_eq}
\end{align}
denotes the tunneling Hamiltonian. Here, $c_{\nu\bm{k\sigma}}$ is the fermionic annihilation operator for system $\nu$ with spin $\sigma$ and momentum $\bm{k}$, $\epsilon_{\bm{k}}$ is the energy dispersion, $U_{R}\:(>0)$ is the pairing interaction strength, $V_{\nu\mu}$ is the amplitude of the tunneling between system $\nu$ and system $\mu$, and $N_{0}$ is the initial particle number of each system. The sum runs over $\langle\nu,\mu\rangle=(1,2), (2,3), (3,1)$. When system 2 is subject to two-body loss with rate $\gamma$, the time evolution of the density matrix $\rho$ is described by the following Lindblad equation~\cite{daley14}
\begin{align}
    \dot{\rho} &= \mathcal{L}\rho = -i[H_\mathrm{sys},\rho] -\frac{\gamma}{2}(\{L^{\dagger}L,\rho\} - 2L\rho L^{\dagger}), \label{tri_liouvillian_eq}
\end{align}
where the two-body loss process is described by the Lindblad operator $L=\sum_{\bm{k}}c_{2-\bm{k}\downarrow}c_{2\bm{k}\uparrow}$, which can be naturally implemented using photoassociation techniques in ultracold atoms~\cite{yamamoto19,yamamoto21,mazza23,nakagawa20,honda23}.
\cred{Since we consider lossy dynamics of fermionic superfluids, the system eventually evolves toward the vacuum state in the long-time limit, irrespective of the dissipation strength and the initial state.
Thus, we focus on the transient dynamics that are accessible in current experimental platforms in ultracold atoms \cite{scott19,yamamoto21}.}

We start by formulating the dissipative BCS theory to describe the nonequilibrium dynamics of multi-terminal Josephson junctions with two-body loss. We first rewrite the Liouvillian $\mathcal{L}$ in Eq.~\eqref{tri_liouvillian_eq} using a ladder representation in the doubled Hilbert space~\cite{shibata19_ashkin,shibata19_kitaev,yamamoto21} as
\begin{align}
    i\mathcal{L}&= H_{\mathrm{sys},{+}} - H_{\mathrm{sys},{-}} +i\gamma(L_{+}L_{-}^{\dagger} - \frac{1}{2}L_{+}^{\dagger}L_{+} - \frac{1}{2}L_{-}^{\dagger}L_{-}) \notag \\
    &=  \tilde{H}_{\mathrm{sys},+}- \tilde{H}_{\mathrm{sys},-}^{\ast} + H_{\text{jump}},
\end{align}
with
\begin{align}
    H_{\text{jump}} &= i\gamma \sum_{\bm{k},\bm{k}'}d_{2\bm{k}\uparrow }^{\dagger}d_{2-\bm{k}\downarrow-}^{\dagger}c_{2-\bm{k}'\downarrow }c_{2\bm{k}'\uparrow}.
\end{align}
Here, we have introduced the subscript $+\:(-)$ for describing the Hamiltonian and the jump operator that act on the ket (bra) space. In the framework of Keldysh path integrals~\cite{sieberer16, yamamoto21}, the subscript $+\:(-)$ corresponds to the forward (backward) path. The jump operators are defined as $L_{+}=\sum_{\bm{k}}c_{2-\bm{k}\downarrow}c_{2\bm{k}\uparrow}$ and $L_{-}=\sum_{\bm{k}}d_{2-\bm{k}\downarrow}d_{2\bm{k}\uparrow}$, where $c_{\nu \bm k\sigma}\:(d_{\nu \bm k\sigma})$ denotes the fermionic annihilation operator that acts on the ket (bra) space. The effective Hamiltonian $\tilde{H}_{\mathrm{sys},\alpha}\:(\alpha=\pm)$ is introduced as
\begin{equation}
    \tilde{H}_{\mathrm{sys},\alpha} = H_{\mathrm{sys},{\alpha}} - \frac{i\gamma}{2}L_{\alpha}^{\dagger}L_{\alpha} =\sum_{\nu}\tilde{H}_{\nu\alpha} +{H}_{\text{tun},  \alpha},
\end{equation}
with 
\begin{align}
\tilde{H}_{\nu+}&= \sum_{\bm{k}\sigma}\epsilon_{\bm{k}}c_{\nu \bm{k} \sigma}^{\dagger}c_{\nu \bm{k} \sigma} -U_{\nu}\sum_{\bm{k},\bm{k}'}c_{\nu\bm{k}\uparrow}^{\dagger}c_{\nu-\bm{k}\downarrow}^{\dagger}c_{\nu-\bm{k}'\downarrow}c_{\nu\bm{k}'\uparrow}, \\
\tilde{H}_{\nu-}&= \sum_{\bm{k}\sigma}\epsilon_{\bm{k}}d_{\nu \bm{k} \sigma}^{\dagger}d_{\nu \bm{k} \sigma} -U_{\nu}\sum_{\bm{k},\bm{k}'}d_{\nu\bm{k}\uparrow}^{\dagger}d_{\nu-\bm{k}\downarrow}^{\dagger}d_{\nu-\bm{k}'\downarrow}d_{\nu\bm{k}'\uparrow},
\end{align}
where $U_\nu$ is defined by $U_{1}=U_{3}=U_{R}$ and $U_{2}=U_{R}+i\gamma/2$. 

Then, we perform the mean-field decoupling for the Liouvillian by introducing the following superfluid order parameters for system $\nu$:
\begin{align}
    &\Delta_{\nu +} = -\frac{U_{\nu}}{N_{0}}\sum_{\bm{k}}\langle c_{\nu-\bm{k}\downarrow}c_{\nu\bm{k}\uparrow}\rangle, \label{triquench_gapeq1} \\ 
    &\bar{\Delta}_{\nu +} = -\frac{U_{\nu}}{N_{0}}\sum_{\bm{k}}\langle c_{\nu\bm{k}\uparrow}^{\dagger}c_{\nu-\bm{k}\downarrow}^{\dagger}\rangle, \label{triquench_order2_p_def} \\
    &\Delta_{\nu -} = -\frac{U_{\nu}^{\ast}}{N_{0}}\sum_{\bm{k}}\langle d_{\nu-\bm{k}\downarrow}d_{\nu\bm{k}\uparrow}\rangle,\\
    &\bar{\Delta}_{\nu -} = -\frac{U_{\nu}^{\ast}}{N_{0}}\sum_{\bm{k}}\langle d_{\nu\bm{k}\uparrow}^{\dagger}d_{\nu-\bm{k}\downarrow}^{\dagger}\rangle,  \label{triquench_gapeq4}
\end{align}
where the expectation value of an operator $A$ is defined as $\langle A\rangle=\text{tr}[A\rho]$. 
First, by applying the mean-field approximation to $\tilde{H}_{\nu\alpha}$, we obtain 
\begin{align}
    &\tilde{H}_{\nu+} = \sum_{\bm{k}\sigma} \epsilon_{\bm{k}}c_{\nu\bm{k}\sigma}^{\dagger}c_{\nu\bm{k}\sigma} \notag \\ 
    &+ \sum_{\bm{k}}(\Delta_{\nu +}c_{\nu\bm{k}\uparrow}^{\dagger}c_{\nu -\bm{k}\downarrow}^{\dagger} + \bar{\Delta}_{\nu +}c_{\nu -\bm{k}\downarrow}c_{\nu\bm{k}\uparrow}), \label{triquench_Hnp_eq} \\
    &\tilde{H}_{\nu-} = \sum_{\bm{k}\sigma} \epsilon_{\bm{k}}d_{\nu\bm{k}\sigma}^{\dagger}d_{\nu\bm{k}\sigma} \notag \\ 
    &+ \sum_{\bm{k}}(\Delta_{\nu -}d_{\nu\bm{k}\uparrow}^{\dagger}d_{\nu -\bm{k}\downarrow}^{\dagger} + \bar{\Delta}_{\nu -}d_{\nu -\bm{k}\downarrow}d_{\nu\bm{k}\uparrow}), \label{triquench_Hnm_eq}
\end{align}
In Eqs.~\eqref{triquench_Hnp_eq} and \eqref{triquench_Hnm_eq}, and in what follows, we ignore the constant term that does not affect the dynamics of the system.
Also, we can conduct the mean-field approximation for the tunneling Hamiltonian $H_{\text{tun}, \alpha}$ as
\begin{align}
    &{H}_{\text{tun}, +} =\sum_{\langle\nu,\mu\rangle}\sum_{\bm{k}}\left[\frac{V_{\nu\mu}\bar{\Delta}_{\nu +}}{U_{\nu}}c_{\mu -\bm{k} \downarrow}c_{\mu \bm{k} \uparrow}+ \frac{V_{\nu\mu}\Delta_{\mu +}}{U_{\mu}} c_{\nu \bm{k} \uparrow}^{\dagger}c_{\nu -\bm{k} \downarrow}^{\dagger} \right.\notag \\
    &\left.+\frac{V_{\nu\mu}\bar{\Delta}_{\mu +}}{U_{\mu}} c_{\nu -\bm{k} \downarrow}c_{\nu \bm{k} \uparrow} +\frac{V_{\nu\mu}\Delta_{\nu +}}{U_{\nu}} c_{\mu \bm{k} \uparrow}^{\dagger}c_{\mu -\bm{k} \downarrow}^{\dagger} \right], \\
    &{H}_{\text{tun}, -}=\sum_{\langle\nu,\mu\rangle}\sum_{\bm{k}}\left[\frac{V_{\nu\mu}\bar{\Delta}_{\nu -}}{U_{\nu}}d_{\mu -\bm{k} \downarrow}d_{\mu \bm{k} \uparrow}+ \frac{V_{\nu\mu}\Delta_{\mu -}}{U_{\mu}} d_{\nu \bm{k} \uparrow}^{\dagger}d_{\nu -\bm{k} \downarrow}^{\dagger}\right. \notag \\
    &\left.+\frac{V_{\nu\mu}\bar{\Delta}_{\mu -}}{U_{\mu}} d_{\nu -\bm{k} \downarrow}d_{\nu \bm{k} \uparrow} +\frac{V_{\nu\mu}\Delta_{\nu -}}{U_{\nu}} d_{\mu \bm{k} \uparrow}^{\dagger}d_{\mu -\bm{k} \downarrow}^{\dagger} \right],
\end{align}
Finally, the quantum jump term is deformed under the mean-field approximation into
\begin{equation}
    H_{\text{jump}} = -\sum_{\bm{k}}\left[\frac{i\gamma \bar{\Delta}_{2 -}}{U_{2}^{\ast}}c_{2 -\bm{k} \downarrow }c_{2 \bm{k} \uparrow } +\frac{i\gamma \Delta_{2 +}}{U_{2}} d_{2 \bm{k}\uparrow }^{\dagger}d_{2 -\bm{k}\downarrow }^{\dagger} \right]. 
\end{equation}
Thus, we obtain the mean-field Liouvillian as
\begin{align}
    &i\mathcal{L} = \sum_{\nu} \sum_{\bm{k}}\left[\Psi_{\nu\bm{k}+}^{\dagger}
    \begin{pmatrix}
        \epsilon_{\bm{k}} & \Delta_{\nu+}^{\prime} \\ \bar{\Delta}_{\nu+}^{\prime} & -\epsilon_{\bm{k}}
    \end{pmatrix}
    \Psi_{\nu\bm{k}+} \right.\notag \\
    &\left.- \Psi_{\nu\bm{k}-}^{\dagger}
    \begin{pmatrix}
        \epsilon_{\bm{k}} & \Delta_{\nu-}^{\prime} \\ \bar{\Delta}_{\nu-}^{\prime} & -\epsilon_{\bm{k}}
    \end{pmatrix}
    \Psi_{\nu\bm{k}-}\right],
\end{align}
where $\Psi_{\nu\bm{k}+}=(c_{\nu\bm{k}\uparrow},c_{\nu-\bm{k}\downarrow}^{\dagger})^{T}$ and $\Psi_{\nu\bm{k}-}=(d_{\nu\bm{k}\uparrow},d_{\nu-\bm{k}\downarrow}^{\dagger})^{T}$ are the Nambu spinors, and we have introduced
\begin{align}
    \Delta_{\nu+}^{\prime} &= 
    \Delta_{\nu +} + \frac{V_{\nu\mu}\Delta_{\mu +}}{U_{\mu}} + \frac{V_{\lambda\nu}\Delta_{\lambda+}}{U_{\lambda}},\notag \\ 
    &[(\nu,\mu,\lambda)=(1,2,3),(2,3,1),(3,1,2)], \\ 
    \bar{\Delta}_{\nu+}^{\prime} &= \begin{cases}
    \bar{\Delta}_{1 +} + \frac{V_{12}\bar{\Delta}_{2 +}}{U_{2}} + \frac{V_{31}\bar{\Delta}_{3+}}{U_{3}},\; (\nu=1), \\
    \bar{\Delta}_{2 +}  + \frac{V_{23}\bar{\Delta}_{3 +}}{U_{3}}+ \frac{V_{12}\bar{\Delta}_{1 +}}{U_{1}} - \frac{i\gamma\bar{\Delta}_{2-}}{U_{2}^{\ast}},\; (\nu=2), \\
    \bar{\Delta}_{3 +} + \frac{V_{31}\bar{\Delta}_{1 +}}{U_{1}} + \frac{V_{23}\bar{\Delta}_{2+}}{U_{2}},\; (\nu=3), \\
    \end{cases}
\end{align}
for the forward path, and 
\begin{align}
    \Delta_{\nu -}^{\prime} &= \begin{cases}
    \Delta_{1 -} + \frac{V_{12}\Delta_{2 -}}{U_{2}^{\ast}} + \frac{V_{31}\Delta_{3-}}{U_{3}}, \quad (\nu=1), \\
    \Delta_{2 -} + \frac{V_{23}\Delta_{3 -}}{U_{3}}+ \frac{V_{12}\Delta_{1 -}}{U_{1}} + \frac{i\gamma\Delta_{2+}}{U_{2}},\quad (\nu=2), \\
    \Delta_{3 -} + \frac{V_{31}\Delta_{1 -}}{U_{1}} + \frac{V_{23}\Delta_{2-}}{U_{2}^{\ast}}, \quad (\nu=3), \\
    \end{cases} \\
    \bar{\Delta}_{\nu-}^{\prime} &= 
    \bar{\Delta}_{\nu -} + \frac{V_{\nu\mu}\bar{\Delta}_{\mu -}}{U_{\mu}^{\ast}} + \frac{V_{\lambda\nu}\bar{\Delta}_{\lambda-}}{U_{\lambda}^{\ast}},\notag \\ 
    &[(\nu,\mu,\lambda)=(1,2,3),(2,3,1),(3,1,2)], 
\end{align}
for the backward path. Importantly, by using the relation $\langle c_{\nu-\bm{k}\downarrow}c_{\nu\bm{k}\uparrow}\rangle =\langle d_{\nu-\bm{k}\downarrow}d_{\nu\bm{k}\uparrow}\rangle = \text{tr}[c_{\nu-\bm{k}\downarrow}c_{\nu\bm{k}\uparrow}\rho]$ and $\text{tr}[A^{\dagger}\rho] = (\text{tr}[A\rho])^{\ast}$, we obtain
\begin{align}
    &\Delta_{\nu +} = \bar{\Delta}_{\nu -}^{\ast},\\
    &\Delta_{\nu -} = \bar{\Delta}_{\nu + }^{\ast},\\
    &\bar{\Delta}_{\nu +} = \frac{\Delta_{\nu +}^{\ast}U_{\nu}}{U_{\nu}^{\ast}},
\end{align}
which lead to 
\begin{equation}
    \Delta_{\nu +}^{\prime} = (\bar{\Delta}_{\nu +}^{\prime})^{\ast} = \Delta_{\nu -}^{\prime} = (\bar{\Delta}_{\nu -}^{\prime})^{\ast}.
\end{equation}
By rewriting the superfluid order parameter $\Delta_{\nu\alpha}\; ({\Delta}_{\nu\alpha}^\prime)$ as $\Delta_{\nu}\; ({\Delta}_{\nu}^\prime)$, we obtain the equation of the time evolution for the density matrix as
\begin{align}
    &\dot{\rho} = -i[H_{\text{eff}},\rho], \label{tri_DMevol_eq} \\ 
    &H_{\text{eff}} = \sum_{\nu\bm{k}} \Psi_{\nu\bm{k}}^{\dagger}
    \begin{pmatrix}
        \epsilon_{\bm{k}} & \Delta_{\nu}^{\prime} \\ \Delta_{\nu}^{\prime *} & -\epsilon_{\bm{k}}
    \end{pmatrix}
    \Psi_{\nu\bm{k}}, \label{tri_eff_BdG_Hami_eq}
\end{align}
with
\begin{align}
    &\Delta_{\nu}^{\prime} = \Delta_{\nu} + \frac{V_{\nu\mu}\Delta_{\mu}}{U_{\mu}} + \frac{V_{\lambda\nu}\Delta_{\lambda}}{U_{\lambda}}, \notag  \\
    &[(\nu,\mu,\lambda)=(1,2,3),(2,3,1),(3,1,2)].
\end{align}

\subsection{Anderson's pseudospin representation}
To simulate the dynamics of superfluid order parameters and associated Josephson currents, it is useful to introduce the Anderson's pseudospin representation~\cite{barankov06_sync,yuzbashyan06,tsuji15,murotani17,yamamoto21}, where the pseudospin for system $\nu$ is defined as
\begin{equation}
    \bm{\sigma}_{\nu \bm{k}} = \frac{1}{2}\Psi_{\nu \bm{k}}^{\dagger}\bm{\tau}\Psi_{\nu \bm{k}},
\end{equation}
where $\bm{\tau}=(\tau_{x},\tau_{y},\tau_{z})$ is the vector of the Pauli matrices. Using the pseudospin representation, we rewrite the effective Hamiltonian~\eqref{tri_eff_BdG_Hami_eq} as
\begin{equation}
    H_{\text{eff}} = 2\sum_{\nu \bm{k}}\bm{b}_{\nu \bm{k}}\cdot \bm{\sigma}_{\nu \bm{k}},
\end{equation}
with
\begin{equation}
    \bm{b}_{\nu \bm{k}} = (\text{Re}{\Delta}_{\nu}^\prime,\; -\text{Im}{\Delta}_{\nu}^\prime,\; \epsilon_{\bm{k}} ).
\end{equation}
Because the pseudospin satisfies the commutation relations $[\sigma_{\bm{k}}^{i},\sigma_{\bm{k}}^{j}] = i\epsilon_{ijk}\sigma_{\bm{k}}^{k}$, Eq.~\eqref{tri_DMevol_eq} is mapped to the Bloch equation of pseudospins, given by
\begin{align}
    \frac{\mathrm{d}\langle \bm{\sigma}_{\nu \bm{k}}\rangle}{\mathrm{d}t} &= 2\bm{b}_{\nu \bm{k}}\times\langle \bm{\sigma}_{\nu \bm{k}}\rangle, \label{tri_bloch_eq}
\end{align}
where the superfluid order parameter is determined self-consistently through
\begin{align}
    &\Delta_{\nu} = |\Delta_{\nu}|e^{i\theta_{\nu}} = -\frac{U_{\nu}}{N_{0}}\sum_{\bm{k}}(\langle \sigma_{\nu\bm{k}}^{x}\rangle -i \langle \sigma_{\nu\bm{k}}^{y}\rangle). \label{tri_gap_eq}
\end{align}
Here, $\theta_\nu$ represents the U(1) phase of the order parameter $\Delta_\nu$.
The time evolution of the particle number $N_\nu$ of system $\nu$ is given as
\begin{align}
    &\frac{1}{N_{0}}\frac{\mathrm{d}N_{1}}{\mathrm{d}t} = -J_{12}+J_{31} ,\label{triquench_dN1dt_eq} \\
    &\frac{1}{N_{0}}\frac{\mathrm{d}N_{2}}{\mathrm{d}t} = -\frac{2\gamma |\Delta_{2}|^{2}}{|U_{2}|^{2}} +J_{12} - J_{23}, \label{triquench_dN2dt_eq} \\
    &\frac{1}{N_{0}}\frac{\mathrm{d}N_{3}}{\mathrm{d}t} = J_{23}-J_{31}, \label{triquench_dN3dt_eq}
\end{align}
where the Josephson current for each junction is calculated as
\begin{align}
    &J_{12} = -I_{12}\sin(\Delta\theta_{12}- \phi), \label{triquench_Ja_eq} \\
    &J_{23} = -I_{23}\sin(\Delta\theta_{23} + \phi), \label{triquench_Jb_eq} \\
    &J_{31} = -I_{31}\sin(\Delta\theta_{31}). \label{triquench_Jc_eq}
\end{align}
Here, $J_{\nu\mu}$ denotes the current from system $\nu$ to system $\mu$,
$I_{\nu\mu} = 4V_{\nu\mu}|\Delta_{\nu}||\Delta_{\mu}|/|U_{\nu}||U_{\mu}|$, $\Delta\theta_{\nu\mu}=\theta_{\nu}-\theta_{\mu}$, and the phase shift $\phi$ is given by $\phi = \tan^{-1}(-\gamma/2U_{R})$. As shown below, Josephson currents exhibit characteristic features governed by the nonequilibrium dynamics of the order parameter.

\section{Nonequilibrium transitions between dynamical phases}\label{sec_DPT}
In this section, we show the numerical results obtained by solving the Bloch equation~\eqref{tri_bloch_eq} together with the order parameter~\eqref{tri_gap_eq}. In the following calculations, the Runge-Kutta method of the $4$th order is used by assuming a constant density of states with $1/W$~\cite{barankov06_sync,yamamoto21}, where $W$ is the bandwidth.
\cred{The bandwidth enters the Lindblad equation only through the noninteracting Hamiltonian part.
Then, if we change the scale of $W$, we need to change the scale of $\gamma$ correspondingly.}
As an initial state, we take $\langle \sigma_{\nu\bm{k}}^{x}\rangle|_{t=0} = -\Delta_{0}/2\sqrt{\epsilon_{\bm{k}}^{2}+\Delta_{0}^{2}}$, $\langle \sigma_{\nu\bm{k}}^{y}\rangle|_{t=0} = 0$, and $\langle \sigma_{\nu\bm{k}}^{z}\rangle|_{t=0} = -\epsilon_{\bm{k}}/2\sqrt{\epsilon_{\bm{k}}^{2}+\Delta_{0}^{2}}$ with $\Delta_{0}\in \mathbb{R}$, corresponding to the BCS ground state. We consider the time evolution after a sudden switch-on of two-body loss $\gamma$ and the tunnelings $V_{12},V_{23}$, and $V_{31}$. The tunneling amplitudes and the interaction strength are set to \cred{$V_{12}=0.02\Delta_{0} ,$ $V_{23}=0.041\Delta_{0}$ and $U_{R}=3.1\Delta_{0}$}. We consider two cases of weak and strong tunnelings for $V_{31}$ since dissipation-induced transitions between dynamical phases exhibit distinct behavior by changing $V_{31}$.  
\cred{We remark that the word “nonequilibrium phase transition” in our study refers to transitions between two dynamical phases with distinct dynamical properties \cite{scott19}, rather than to transitions characterized by nonanalyticities in the fidelity at finite times~\cite{Heyl18}.}

\subsection{\cred{Two-step nonequilibrium phase transition} for weak $V_{31}$}\label{subsec_V31weak}
\begin{figure}[b]
    \centering
    \includegraphics[width=\linewidth]{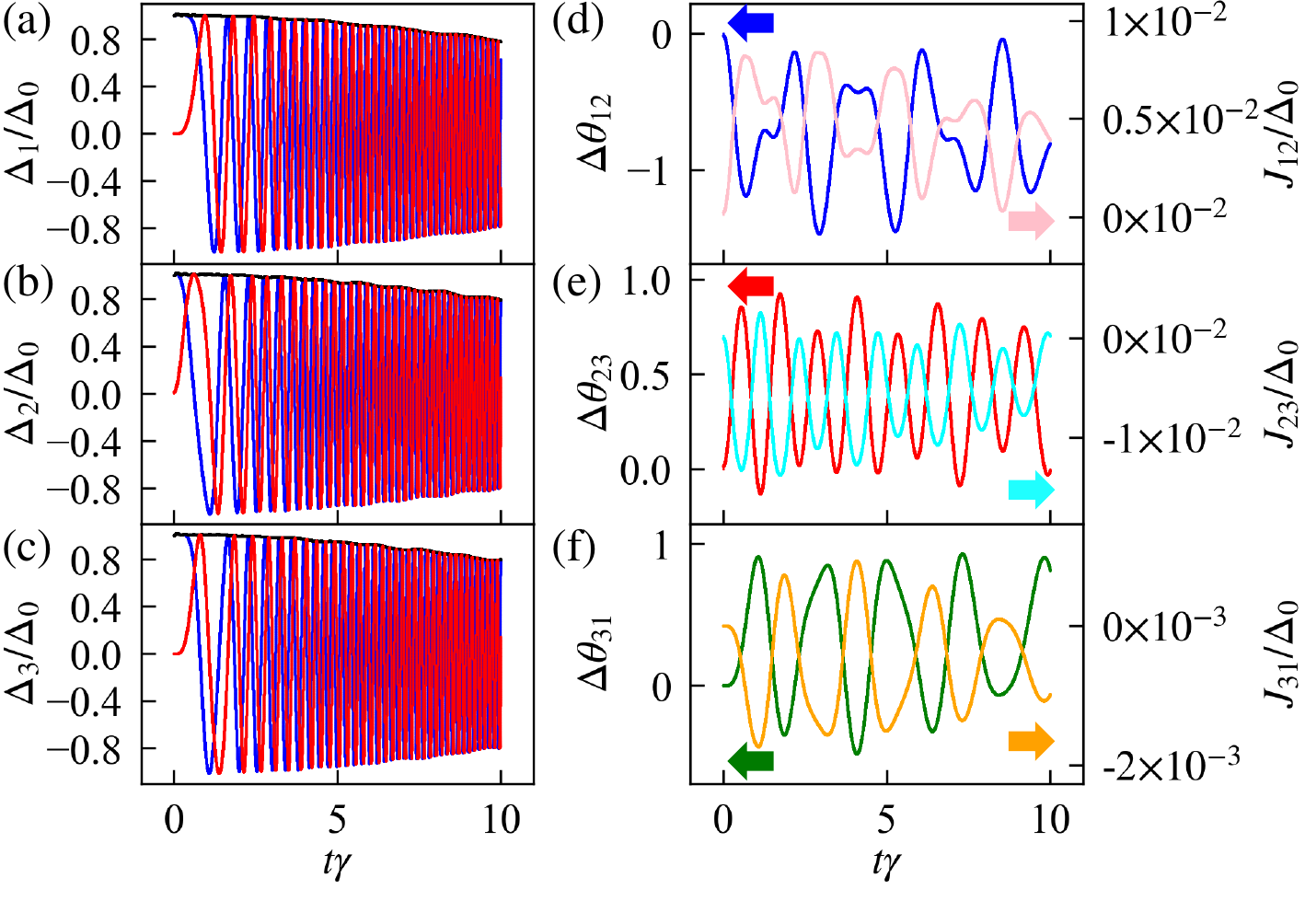}
    \caption{(a), (b), (c) Nonequilibrium dynamics of order parameters $\Delta_{1},\Delta_{2}$ and $\Delta_{3}$ following a sudden switch-on of two-body loss. (d), (e), (f) Phase differences $\Delta\theta_{12}$, $\Delta\theta_{23}$, and $\Delta\theta_{31}$, and corresponding Josephson currents $J_{12}$, $J_{23},$ and $J_{31}$ are plotted. In (a), (b), and (c), the real part (blue), the imaginary part (red), and the amplitude (black) of order parameters are shown.
    The case of weak loss \cred{$\gamma=0.082\Delta_{0}$} is displayed. All Josephson currents have finite dc components. The parameters are set to \cred{$\:V_{31}=0.0051\Delta_{0},\:V_{12}=0.02\Delta_{0} ,\:V_{23}=0.041\Delta_{0},\:U_{R}=3.1\Delta_{0},$ and $W=5.1\Delta_{0}$}.}
    \label{V310.05g0.8_image}
\end{figure}
First, we consider the case where the tunneling $V_{31}$ between system 1 and system 3 is weak (\cred{$V_{31}=0.0051\Delta_{0}$}). Figures~\ref{V310.05g0.8_image}(a)-(c) show the dynamics of the order parameters when the loss is weak (\cred{$\gamma=0.082\Delta_{0}$}). Note that we find qualitatively similar behavior for \cred{$\gamma \le 0.1\Delta_{0}$ (see Appendix~\ref{app_weak})}. An important point is that the sudden switch-on of two-body loss causes the chirped phase rotation of the order parameter, because of the decrease of the particle number of the system. Moreover, as shown in Figs.~\ref{V310.05g0.8_image}(d)-(f), the relative phase $\Delta\theta_{\nu\mu}$ of order parameters shows an oscillation hosting relatively small values shifted by a positive constant, which means that the dynamics of the order parameter manifests synchronization. Correspondingly, Josephson currents $J_{12},J_{23}$ and $J_{31}$ in Figs.~\ref{V310.05g0.8_image}(d)-(f) have both dc and ac components. Here, the dc component of the Josephson current $J_{\nu\mu}$ is defined as $J_{\nu\mu}^\text{dc} =1/(t_f-t_i) \int_{t_{i}}^{t_{f}}\mathrm{d}t J_{\nu\mu}(t)$ with $t_{i}\gamma=0.5$ and $t_{f}\gamma=10$. We note that the phase difference $\Delta\theta_{\nu\mu}$ displays an oscillation resembling the superposition of two sinusoidal waves due to the condition $V_{12}\neq V_{23}$ for weak $\gamma$, thereby Josephson currents exhibit a modulation pattern arising from two different frequencies (see Appendix~\ref{app_beating}).

\begin{figure}[t]
    \centering
    \includegraphics[width=\linewidth]{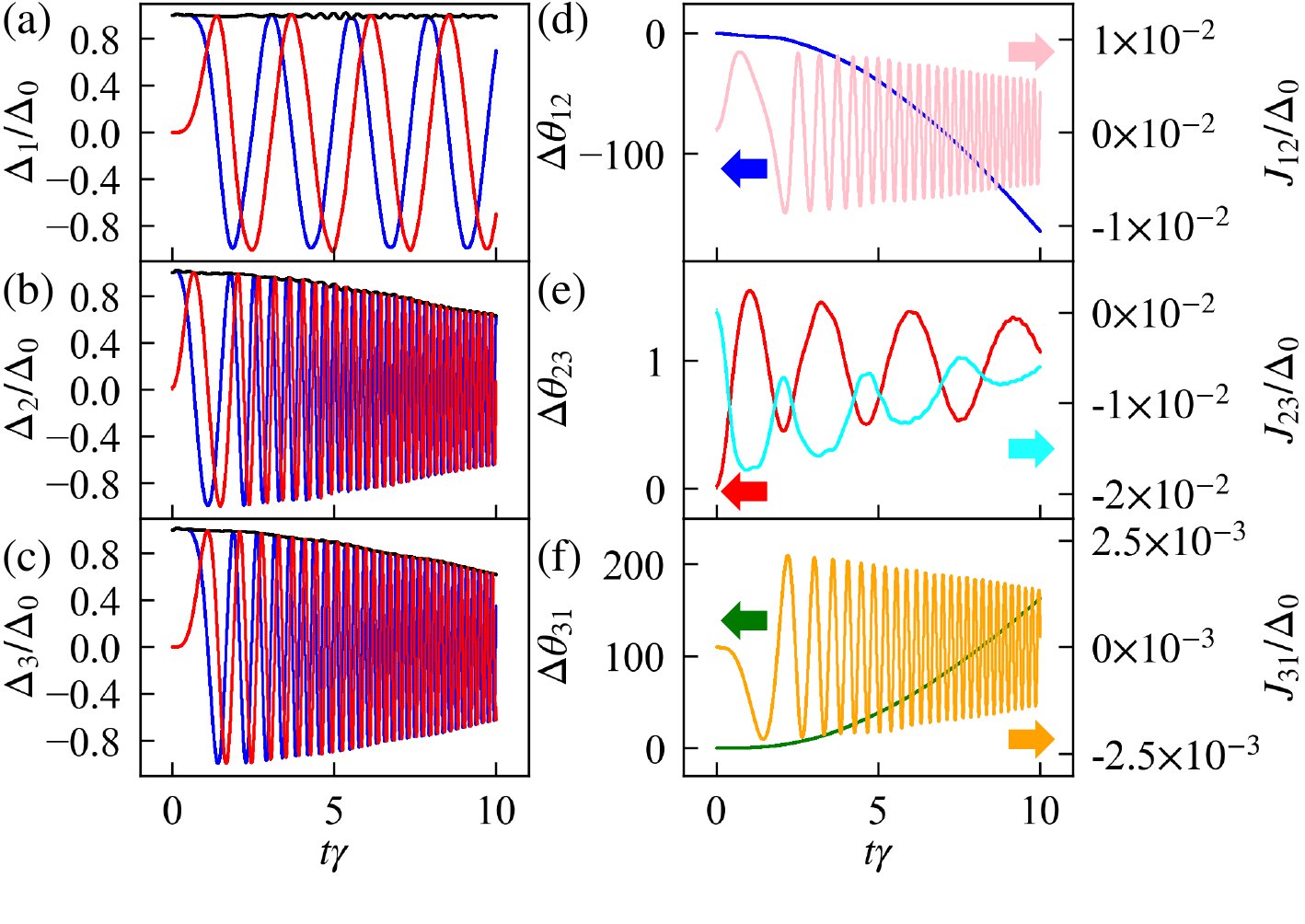}
    \caption{(a), (b), (c) Nonequilibrium dynamics of order parameters $\Delta_{1},\Delta_{2}$ and $\Delta_{3}$ following a sudden switch-on of two-body loss. (d), (e), (f) Phase differences $\Delta\theta_{12}$, $\Delta\theta_{23}$, and $\Delta\theta_{31}$, and corresponding Josephson currents $J_{12}$, $J_{23},$ and $J_{31}$ are plotted. In (a), (b), and (c), the real part (blue), the imaginary part (red), and the amplitude (black) of order parameters are shown.
    The case of medium loss \cred{$\gamma=0.13\Delta_{0}$} is displayed. Only $J_{23}$ has a finite dc component. The other parameters are the same as those in Fig.~\ref{V310.05g0.8_image}.}
    \label{V310.05g1.11_image}
\end{figure}

Then, for medium dissipation (\cred{$\gamma=0.13\Delta_{0}$}) as shown in Figs.~\ref{V310.05g1.11_image}(a)-(c), we find that the order parameters of system 2 and system 3 show faster oscillations and decays than those of system 1. Consequently, the synchronization of order parameters occurs only between system 2 and system 3. This is apparent from the dynamics of the phase difference $\Delta\theta_{\nu\mu}$ shown in Figs.~\ref{V310.05g1.11_image}(d)-(f), where $|\Delta\theta_{12}|$ and $|\Delta\theta_{31}|$ rapidly grow over time, while the phase difference $\Delta\theta_{23}$ oscillates hosting small amplitudes over time. Correspondingly, in Figs.~\ref{V310.05g1.11_image}(d) and \ref{V310.05g1.11_image}(f), the Josephson currents $J_{12}$ and $J_{31}$ only exhibit ac components; in contrast, in Fig.~\ref{V310.05g1.11_image}(e), the Josephson current $J_{23}$ has a nonzero dc component together with the ac component. This means that the \cred{nonequilibrium phase transition} characterized by the vanishing dc components of $J_{12}$ and $J_{31}$ emerges with increasing two-body loss $\gamma$. We note that the critical dissipation strength is estimated as \cred{$0.1\Delta_{0}< \gamma_{c_{1}}< 0.12\Delta_{0}$}, but we find that the numerical simulation is unstable around the transition point, and it is rather hard to determine the precise value of $\gamma_{c_{1}}$.
Moreover, the vanishing dc Josephson current indicates the suppression of inflow or outflow of particles via the junction (see Appendix~\ref{app_particle}). We find that, since $J_{12}$ and $J_{31}$ only have ac components, the particle number of system 1 decays very slowly, the fact of which is physically reminiscent of the continuous quantum Zeno effect~\cite{yamamoto21,garcia09,daley09,yamamoto19,syassen08,yan13,zhu14}. 
In contrast, the order parameters in system 2 and system 3 both show damped oscillations with almost equivalent decay rates, leading to a net particle flow from system 3 to system 2.
\begin{figure}[t]
    \centering
    \includegraphics[width=\linewidth]{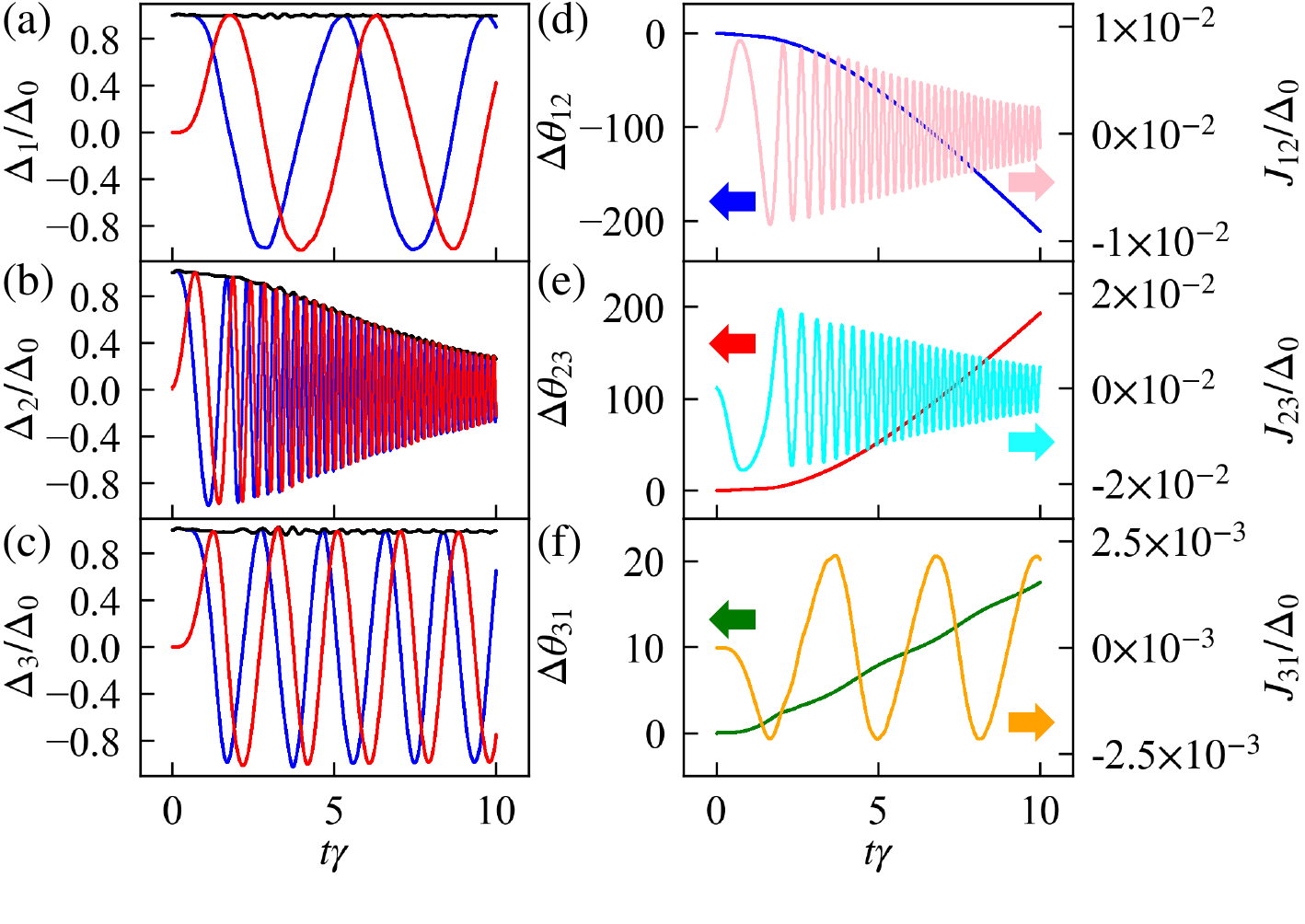}
    \caption{(a), (b), (c) Nonequilibrium dynamics of order parameters $\Delta_{1},\Delta_{2}$ and $\Delta_{3}$ following a sudden switch-on of two-body loss. (d), (e), (f) Phase differences $\Delta\theta_{12}$, $\Delta\theta_{23}$, and $\Delta\theta_{31}$, and corresponding Josephson currents $J_{12}$, $J_{23},$ and $J_{31}$ are plotted. In (a), (b), and (c), the real part (blue), the imaginary part (red), and the amplitude (black) of order parameters are shown. All Josephson currents do not have finite dc components.
    The case of strong loss \cred{$\gamma=0.15\Delta_{0}$} is displayed. The other parameters are the same as those in Fig.~\ref{V310.05g0.8_image}.}
    \label{V310.05g1.5_image}
\end{figure}

Finally, for strong dissipation (\cred{$\gamma=0.15\Delta_{0}$}), the order parameters of system 1 and system 3 exhibit oscillations slower than that of system 2, the latter of which rapidly decays over time [see Figs.~\ref{V310.05g1.5_image}(a)-(c)]. We note that the order parameters show qualitatively similar behavior for \cred{$\gamma>0.14\Delta_{0}$}. This leads to an increase in all the absolute values of the phase differences $|\Delta\theta_{\nu\mu}|$ as shown in Figs.~\ref{V310.05g1.5_image}(d)-(f), and all dc components of the Josephson currents $J_{\nu\mu}$ vanish. This implies that the second \cred{nonequilibrium phase transition} characterized by the vanishing all dc Josephson currents occurs with increasing $\gamma$. Correspondingly, the particle numbers of system 1 and system 3 decay very slowly, which is again reminiscent of the continuous quantum Zeno effect (see Appendix~\ref{app_particle}). In our calculation, the critical point for this second \cred{nonequilibrium phase transition} occurs at \cred{$\gamma_{c_{2}}\simeq 0.14\Delta_{0}$}. We note that, for arbitrary dissipation strengths, the ac Josephson currents in all junctions persist over time. \cred{We also confirm that our results remain qualitatively unchanged for different values of $U_{R}$ and $V_{\nu\mu}$ (see Appendix~\ref{app_UV}).}

\subsection{\cred{Nonequilibrium phase transition} for strong $V_{31}$}
\begin{figure}[t]
    \centering
    \includegraphics[width=8.57cm]{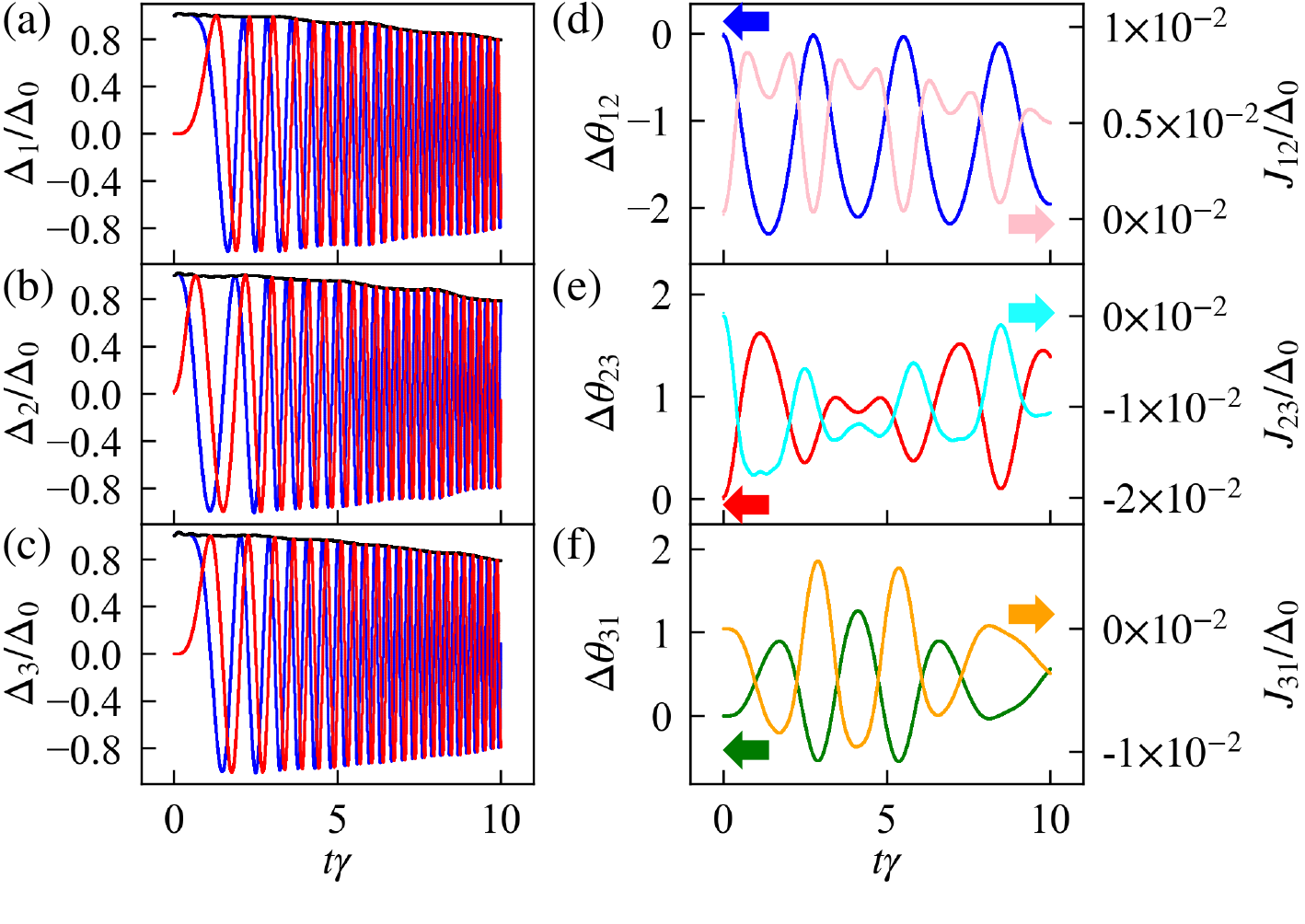}
    \caption{(a), (b), (c) Nonequilibrium dynamics of order parameters $\Delta_{1},\Delta_{2}$ and $\Delta_{3}$ following a sudden switch-on of two-body loss. (d), (e), (f) Phase differences $\Delta\theta_{12}$, $\Delta\theta_{23}$, and $\Delta\theta_{31}$, and corresponding Josephson currents $J_{12}$, $J_{23},$ and $J_{31}$ are plotted. In (a), (b), and (c), the real part (blue), the imaginary part (red), and the amplitude (black) of order parameters are shown. All Josephson currents have finite dc components.
    The case of medium loss \cred{$\gamma=0.13\Delta_{0}$} is displayed and the tunneling amplitude is set to \cred{$V_{31}=0.026\Delta_{0}$}. The other parameters are the same as those in Fig.~\ref{V310.05g0.8_image}.}
    \label{V310.25g1.11_image}
\end{figure}

Next, we consider the case when the tunneling amplitude $V_{31}$ is strong (\cred{$V_{31}=0.026\Delta_{0}$}). In this case, the dynamics of the order parameter exhibits distinct behavior compared to the case for weak $V_{31}$. In particular, the dynamics of physical quantities exhibits qualitatively similar behavior for weak and medium dissipation strengths. In Figs.~\ref{V310.25g1.11_image}(a)-(f), we show the dynamics of order parameters and associated phase difference together with the Josephson current for $\cred{\gamma=0.13\Delta_{0}}$. We find that, even when $\gamma$ is of medium strength, order parameters exhibit synchronization characterized by the finite dc and ac components of all the Josephson currents. In addition, we see a beating oscillation of the phase difference as shown in Figs.~\ref{V310.25g1.11_image}(d)-(f) (see Appendix~\ref{app_beating}). We note that, when $\cred{\gamma\le0.13\Delta_{0}}$, the dynamics of order parameters and associated phase differences together with the Josephson currents are qualitatively the same as in Fig.~\ref{V310.25g1.11_image}.

\begin{figure}[t]
    \centering
    \includegraphics[width=8.57cm]{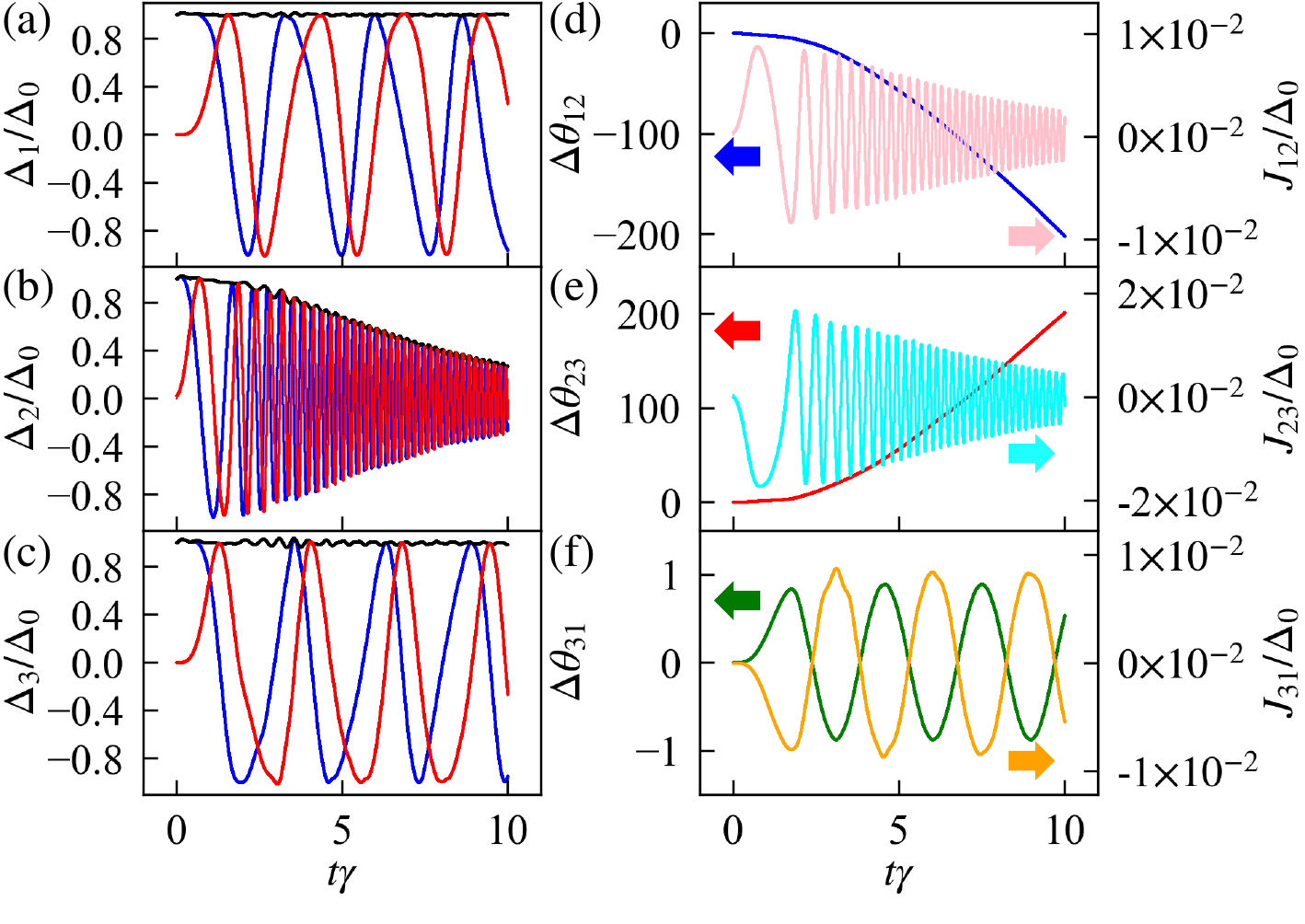}
    \caption{(a), (b), (c) Nonequilibrium dynamics of order parameters $\Delta_{1}$, $\Delta_{2}$, and $\Delta_{3}$ following a sudden switch-on of two-body loss. (d), (e), (f) Phase differences $\Delta\theta_{12}$, $\Delta\theta_{23}$, and $\Delta\theta_{31}$, and corresponding Josephson currents $J_{12}$, $J_{23},$ and $J_{31}$ are plotted. In (a), (b), and (c), the real part (blue), the imaginary part (red), and the amplitude (black) of order parameters are shown. All Josephson currents do not have finite dc components.
    The case of strong loss \cred{$\gamma=0.15\Delta_{0}$} is displayed. The other parameters are the same as those in Fig.~\ref{V310.25g1.11_image}.}
    \label{V310.25g1.5_image}
\end{figure}

Remarkably, in the presence of strong dissipation (\cred{$\gamma=0.15\Delta_{0}$}) as shown in Fig.~\ref{V310.25g1.5_image}, all dc Josephson currents simultaneously vanish. We note that qualitatively the same dynamics is observed when \cred{$\gamma\ge0.14\Delta_{0}$}. This demonstrates that the \cred{nonequilibrium phase transition} arises characterized by the simultaneous disappearance of dc Josephson currents in all junctions, which is in stark contrast to the case of the two-step \cred{nonequilibrium phase transition} for weak $V_{31}$ discussed in Sec.~\ref{subsec_V31weak}. Physically, this distinct behavior between weak and strong $V_{31}$ is explained as follows. When $V_{31}$ is strong, system 1 and system 3 effectively form a single fermionic superfluid due to the strong connection between them. Then, the dynamics of the system is regarded as that of two dissipative fermionic superfluids connected via Josephson junctions, and thereby \cred{the nonequilibrium phase transition} occurs only once. We remark that, though the relative phase $\Delta\theta_{31}$ shows oscillations hosting small values, the center of the oscillation is at $\Delta\theta_{31}=0$, and the dc Josephson current does not appear. This mechanism is distinct from the case of weak $V_{31}$ discussed in Fig.~\ref{V310.05g1.5_image}. This ac oscillation for $J_{31}$ resembles the behavior of the conventional ac Josephson current~\cite{valtolina15,burchianti18,luick20,kwon20,del21,del25}. In the numerical calculation, \cred{the nonequilibrium phase transition} occurs only once when \cred{$V_{31}\ge0.02\Delta_{0}$}, and the critical dissipation strength is estimated at \cred{$0.13\Delta_{0}<\gamma_{c_{2}}<0.14\Delta_{0}$}. Interestingly, we find that the transition point $\gamma_{c_{2}}$ seems to show almost no shift even when $V_{31}$ is increased. Therefore, by tuning the tunneling $V_{31}$, two distinct types of \cred{nonequilibrium phase transitions} arise induced by the interplay between two-body loss and the multiple coherence via Josephson junctions.

\section{Analysis based on a simplified model}\label{sec_Simp}
In this section, by using Eqs.~\eqref{triquench_dN1dt_eq}-\eqref{triquench_dN3dt_eq}, we derive simplified equations for the dynamics of Josephson currents and analytically explain the \cred{nonequilibrium phase transition} obtained in Sec.~\ref{sec_DPT}. 
\cred{Specifically, we treat the case of $V_{31} = 0$ and evaluate the equations, demonstrating that nonequilibrium phases are classified into three types and that analytical results qualitatively reproduce those observed
by solving the Bloch equation~\eqref{tri_bloch_eq}.}
\begin{figure*}[bth]
    \centering
    \includegraphics[width=\textwidth]{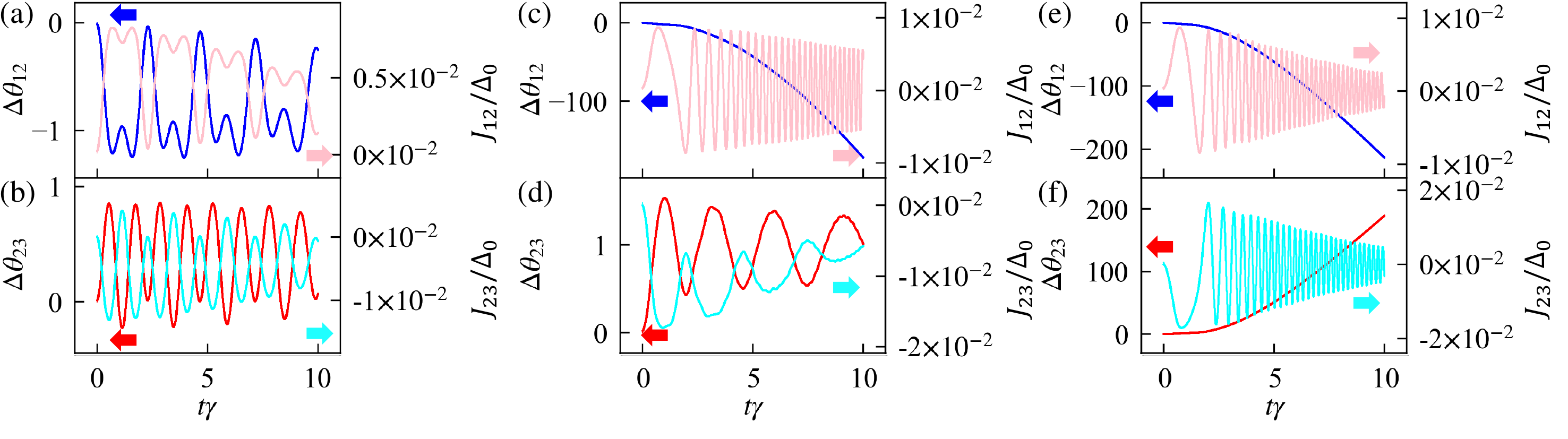}
    \caption{Nonequilibrium dynamics of the phase differences $\Delta\theta_{12}$ and $\Delta\theta_{23}$, and Josephson currents $J_{12}$ and $J_{23}$, obtained by solving the Bloch equation~\eqref{tri_bloch_eq} following a sudden switch-on of two-body loss for the BCS ground state. The dissipation strength is set to (a), (b) \cred{$\gamma=0.082\Delta_{0}$}, (c), (d) \cred{$\gamma=0.13\Delta_{0}$}, and (e), (f) \cred{$\gamma=0.15\Delta_{0}$} corresponding to weak, medium, and strong dissipation that were used in Figs.~\ref{V310.05g0.8_image}, ~\ref{V310.05g1.11_image}, and ~\ref{V310.05g1.5_image}, respectively.
    For weak dissipation [(a) and (b)], all Josephson currents have finite dc components. For medium dissipation [(c) and (d)], only $J_{23}$ has a finite dc component. For strong dissipation [(e) and (f)], all Josephson currents have no dc components. The other parameters are the same as those in Fig.~\ref{V310.05g0.8_image} except for $V_{31}=0$.}
    \label{V310_image}
\end{figure*}

\subsection{Numerical results for $V_{31}=0$ using the Bloch equation \eqref{tri_bloch_eq}}\label{subsec_simple_numeric}
First, in Fig.~\ref{V310_image}, we show the phase differences and corresponding Josephson currents obtained by solving the Bloch equation~\eqref{tri_bloch_eq} for $V_{31}=0$. We note that, because the tunneling satisfies $V_{31}=0$, there exists no current between system 1 and system 3. We find that, for $V_{31}=0$, the results are similar to those for weak $V_{31}$ obtained in Figs.~\ref{V310.05g0.8_image}-\ref{V310.05g1.5_image}. For weak dissipation $\gamma$ [Figs.~\ref{V310_image}(a) and \ref{V310_image}(b)], the phase differences $\Delta\theta_{12}$ and $\Delta\theta_{23}$ exhibit an oscillation shifted by a constant from the origin, and corresponding Josephson currents $J_{12}$ and $J_{23}$ show nonzero dc components. For medium dissipation [Figs.~\ref{V310_image}(c) and \ref{V310_image}(d)], $|\Delta\theta_{12}|$ rapidly grows over time while $\Delta\theta_{23}$ still oscillates with time, resulting in the disappearance of the dc Josephson current $J_{12}^\text{dc}$. For strong dissipation [Figs.~\ref{V310_image}(e) and \ref{V310_image}(f)], the absolute values of the phase differences increase over time and all dc components of Josephson currents vanish. In the numerical calculation, \cred{the two-step nonequilibrium phase transition} occurs, and the critical dissipation strength is estimated at \cred{$0.098\Delta_{0} < \gamma_{c_{1}} < 0.12\Delta_{0}$ and $\gamma_{c_{2}}\simeq 0.14\Delta_{0}$.}

\subsection{Analytical results using a simplified model}
Here, we derive the equation of motion for the phase difference. Using Eqs.~\eqref{triquench_dN1dt_eq}-\eqref{triquench_dN3dt_eq} and assuming the following relation between the particle number and the phase difference \cite{yamamoto21}:
\begin{align}
    &\frac{\mathrm{d}\Delta\theta_{12}}{\mathrm{d}t} = -\frac{W}{N_{0}}(N_{1}-N_{2}), \\
    &\frac{\mathrm{d}\Delta\theta_{23}}{\mathrm{d}t} = -\frac{W}{N_{0}}(N_{2}-N_{3}),
\end{align}
we obtain
\begin{align}
    &\frac{\mathrm{d}^{2}\Delta\theta_{12}}{\mathrm{d}t^{2}} = -W[2I_{12}\sin(\Delta \theta_{12}) -I_{23}\sin(\Delta\theta_{23}) \notag \\
    &- I_{31}\sin(-\Delta\theta_{12}-\Delta\theta_{23}) +C_{0}], \label{TriQuench_phase_a_diff_eq} \\
    &\frac{\mathrm{d}^{2}\Delta\theta_{23}}{\mathrm{d}t^{2}} = -W[2I_{23}\sin(\Delta \theta_{23}) -I_{12}\sin(\Delta\theta_{12}) \notag \\
    &- I_{31}\sin(-\Delta\theta_{12}-\Delta\theta_{23}) -C_{0}], \label{TriQuench_phase_b_diff_eq}
\end{align}
where $C_{0}=2\gamma|\Delta_{2}|^{2}/|U_{2}|^{2}$. Here, we have ignored the phase shift $\phi$ in Eqs.~\eqref{TriQuench_phase_a_diff_eq}-~\eqref{TriQuench_phase_b_diff_eq} because $\phi$ is much smaller than the phase difference and does not affect the results.
Equations~\eqref{TriQuench_phase_a_diff_eq} and \eqref{TriQuench_phase_b_diff_eq} are rewritten as
\begin{align}
    &\frac{\mathrm{d}^{2}}{\mathrm{d}t^{2}} \bm{x}
    = -\frac{\mathrm{d}}{\mathrm{d}\bm{x}}U(x,y), \label{TriQuench_RCSJ_EOM_eq}
\end{align}
with 
\begin{align}
    &U(x,y) = -2W\left[I_{12}\cos (\frac{\sqrt{3}x+y}{2}) + I_{23}\cos(\frac{-\sqrt{3}x+y}{2})\right. \notag \\
    &\left.-I_{31}\cos(y) -\frac{1}{\sqrt{3}}C_{0}x\right], \label{TriQuench_RCSJ_pot_eq}
\end{align}
where $\bm{x}=(x,y)$, $x=(\Delta\theta_{12}-\Delta\theta_{23})/\sqrt{3}$, 
and $y = \Delta\theta_{12}+\Delta\theta_{23}$~\cite{strenski85,geigenmuller96}. Importantly, Eq.~\eqref{TriQuench_RCSJ_EOM_eq} indicates that the time evolution of the phase difference is equivalent to the equation of motion of a particle moving in a washboard potential $U(x,y)$~\cite{caldeira81,schmid83,guinea85}. 

In the following, we consider $V_{12}<V_{23}$ and $V_{31}=0$ to satisfy the conditions given in Sec.~\ref{subsec_simple_numeric}, and demonstrate that three types of particle motions emerge by assuming $|\Delta_{1}|\simeq|\Delta_{2}|\simeq|\Delta_{3}|=|\Delta_{1}|_{t=0}$: (i) the particle is confined in both $\Delta\theta_{12}$ and $\Delta\theta_{23}$ directions, (ii) the particle is confined only in the $\Delta\theta_{23}$ direction, and (iii) the particle is not confined in either direction. These three phases correspond to distinct dynamical phases, and the transitions between them cause the \cred{nonequilibrium phase transition} obtained in Fig.~\ref{V310_image}.

\begin{figure*}[t]
    \centering
    \includegraphics[width=\textwidth]{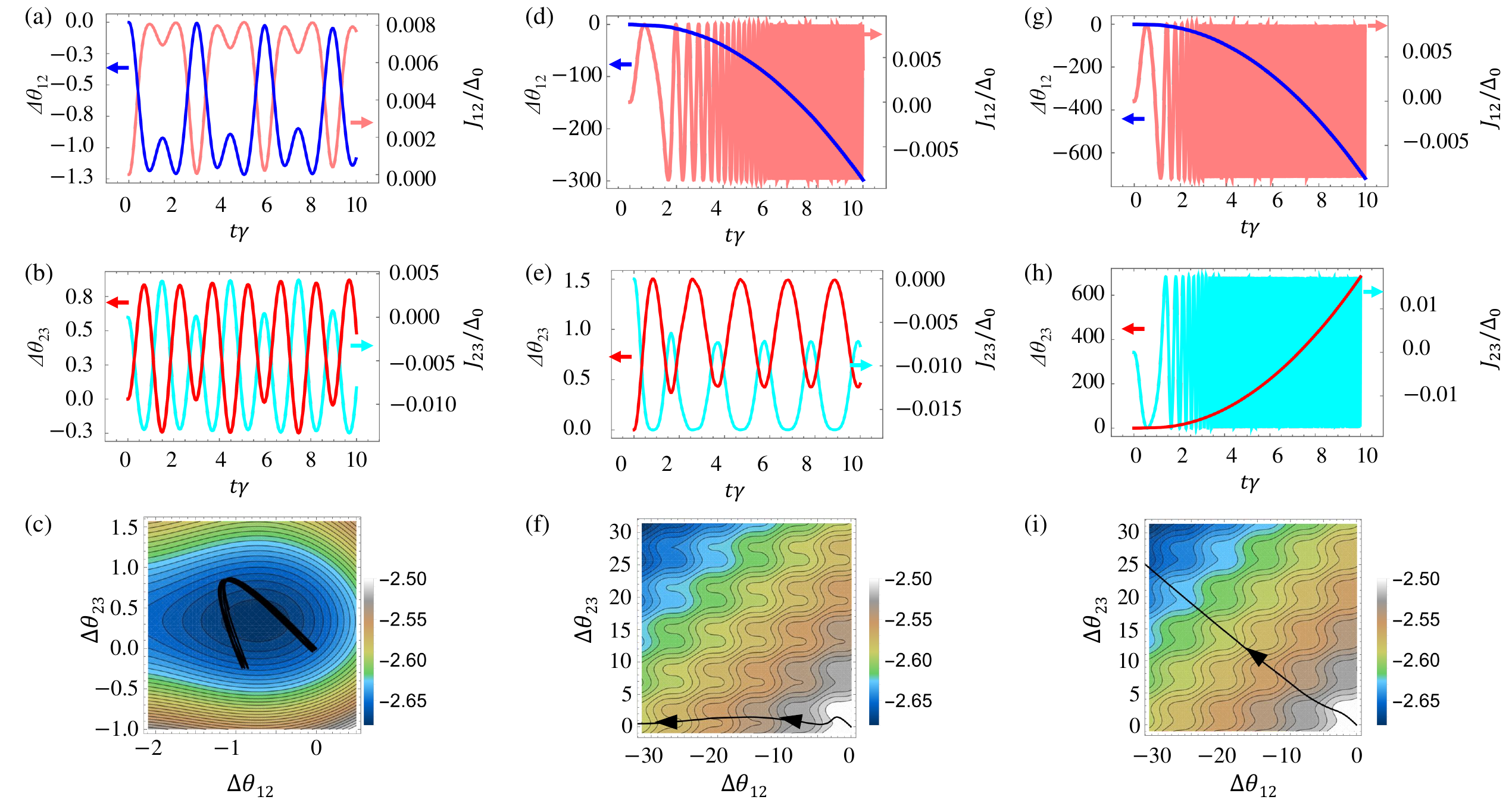}
    \caption{(a), (b), (c) Nonequilibrium dynamics of phase differences $\Delta\theta_{12}$ and $\Delta\theta_{23}$, Josephson currents $J_{12}$ and $J_{23}$, and a trajectory of the particle's motion obtained by solving the equation of motion~\eqref{TriQuench_RCSJ_EOM_eq} under the assumption $|\Delta_{1}|\simeq|\Delta_{2}|\simeq |\Delta_{3}|=|\Delta_{1}|_{t=0}$ following a sudden switch-on of two-body loss for the BCS ground state. The dissipation strength is set to (a)-(c) \cred{$\gamma=0.082\Delta_{0}$}, (d)-(f) \cred{$\gamma=0.13\Delta_{0}$}, and (g)-(i) \cred{$\gamma=0.15\Delta_{0}$} corresponding to weak, medium, and strong dissipation used in Fig.~\ref{V310_image}. The parameters are the same as those in Fig.~\ref{V310_image}. In (c), the contour plot shows the value of the potential $U(x,y)$. For weak dissipation [(a) and (b)], all Josephson currents have finite dc components. For medium dissipation [(d) and (e)], only $J_{23}$ has a finite dc component. For strong dissipation [(g) and (h)], all Josephson currents have no dc components.
    }
    \label{washboard_image}
\end{figure*}

First, so that the particle should be confined in both $\Delta\theta_{12}$ and $\Delta\theta_{23}$ directions, the existence of the minimum of the potential $U(x,y)$ is required. Then, the derivatives of the potential $U(x,y)$ should satisfy $\partial U/\partial x=0$ and $\partial U/\partial y=0$, and the Hessian given by
\begin{equation}
    M = 3W^{2}I_{12}I_{23}\cos(\frac{\sqrt{3}x+y}{2})\cos(\frac{-\sqrt{3}x+y}{2}), \label{triquench_Hesse}
\end{equation}
should be positive. The conditions $\partial U/\partial x=0$ and $\partial U/\partial y=0$ are equivalent to
\begin{align}
    \sin(\frac{\sqrt{3}x+y}{2})=-\frac{C_{0}}{3I_{12}}, \label{triquench_wash_eq1} \\
    \sin(\frac{-\sqrt{3}x+y}{2})=\frac{C_{0}}{3I_{23}}. \label{triquench_wash_eq2}
\end{align}
By substituting the solution that satisfies Eqs.~\eqref{triquench_wash_eq1} and ~\eqref{triquench_wash_eq2} into Eq.~\eqref{triquench_Hesse}, we can confirm that $M>0$ around $(\Delta\theta_{12},\Delta\theta_{23})=(0,0)$, indicating the existence of a minimum of potential when $C_{0}/3I_{12}\le1$ and $C_{0}/3I_{23}\le 1$. Conversely, when $C_{0}/3I_{12}>1$ and $C_{0}/3I_{23}>1$, there is no minimum in the potential $U (x, y)$. By assuming that the dynamics is slow so that $|\Delta_{1}|\simeq |\Delta_{2}|\simeq |\Delta_{3}|=|\Delta_{1}|_{t=0}$, we obtain the critical dissipation strength for the first \cred{nonequilibrium phase transition} with weak dissipation as $\gamma_{c_{1}}=6V_{12}$, which qualitatively agrees well with the numerical results obtained by the Bloch equation~\eqref{tri_bloch_eq} in Fig.~\ref{V310_image} (see Sec.~\ref{subsec_simple_numeric}). Indeed, when $\gamma<\gamma_{c_{1}}$, the solution where the particle is confined in both $\Delta\theta_{12}$ and $\Delta\theta_{23}$ directions can exist, corresponding to the case where the dc Josephson currents for $J_{12}$ and $J_{23}$ do not vanish. When $\gamma>\gamma_{c_{1}}$, the particle is not confined in $\Delta\theta_{12}$ direction, but confined in $\Delta\theta_{23}$ direction, which corresponds to the case where the dc Josephson current for $J_{23}$ survives. We remark that a solution confined only in $\Delta\theta_{12}$ direction does not exist.

Then, the condition required for the particle to be confined only in $\Delta\theta_{23}$ direction is given as
\begin{equation}
    \frac{\partial U(x,y)}{\partial (-\sqrt{3}x+y)}=0.
\end{equation}
Using this condition and assuming that $|\Delta_{1}|\simeq |\Delta_{2}|\simeq |\Delta_{3}|=|\Delta_{1}|_{t=0}$ for simplicity, we obtain $\gamma_{c_{2}}=6V_{23}$, which qualitatively agrees with the results obtained by solving the Bloch equation~\eqref{tri_bloch_eq} in Fig.~\ref{V310_image} (see Sec.~\ref{subsec_simple_numeric}).
When $\gamma>\gamma_{c_{2}}$, the particle cannot be confined in either $\Delta\theta_{12}$ and $\Delta\theta_{23}$ direction. This condition corresponds to the case where all dc Josephson currents vanish. Since $V_{12}<V_{23}$, we have $\gamma_{c_{1}}<\gamma_{c_{2}}$, indicating that the \cred{nonequilibrium phase transition} occurs twice.

Finally, we numerically compare the results obtained by solving the Bloch equation~\eqref{tri_bloch_eq} in Fig.~\ref{V310_image} with those obtained by solving the simplified equation~\eqref{TriQuench_RCSJ_EOM_eq}, assuming that the absolute value of the order parameter is constant as $|\Delta_{1}|=|\Delta_{2}|=|\Delta_{3}|=|\Delta_{1}|_{t=0}$. We consider the time evolution after a sudden switch-on of the two-body loss $\gamma$ for the BCS ground state. We use the same condition as in Fig.~\ref{V310.05g0.8_image}. In Fig.~\ref{washboard_image}, we present the results obtained by solving Eq.~\eqref{TriQuench_RCSJ_EOM_eq}. For weak dissipation shown in Figs.~\ref{washboard_image}(a)-(c), the phase differences $\Delta\theta_{12}$ and $\Delta\theta_{23}$ oscillate displaying a phase modulation, corresponding to the condition where the particle is confined around $(\Delta\theta_{12},\Delta\theta_{23})=(0,0)$. This is equivalent to the case where all the dc Josephson currents for $J_{12}$ and $J_{23}$ remain finite. 
In the case of medium dissipation shown in Figs.~\ref{washboard_image}(d)-(f), the particle is not confined along the $\Delta\theta_{12}$ direction, while being confined along $\Delta\theta_{23}$ direction. This corresponds to the situation where the dc current for $J_{23}$ only becomes finite. For strong dissipation shown in Figs.~\ref{washboard_image}(g)-(i), the particle is unconfined in any directions, which corresponds to the case where the dc components of both $J_{12}$ and $J_{23}$ vanish. Thus, \cred{the two-step nonequilibrium phase transition} characterized by the vanishing dc Josephson currents occurs as the dissipation is increased. We have confirmed that qualitatively the same results are obtained for other $V_{31}$ used in Figs.~\ref{V310.05g0.8_image}-\ref{V310.25g1.5_image} by using Eq.~\eqref{TriQuench_RCSJ_EOM_eq} (not shown). Therefore, the simplified equation of motion~\eqref{TriQuench_RCSJ_EOM_eq} under the assumption $|\Delta_{1}|\simeq|\Delta_{2}|\simeq |\Delta_{3}|=|\Delta_{1}|_{t=0}$ qualitatively captures the same behavior as the solution of the Bloch equation~\eqref{tri_bloch_eq}. 

\cred{We have analytically demonstrated that the dynamical phases can be classified into three types. Since finite $V_{31}$ couples system 1 and system 3, thereby forming an effective single superfluid, the number of transitions can decrease but never increase for finite $V_{31}$.}
Overall, \cred{the nonequilibrium phase transition} of dissipative fermionic superfluids coupled via multi-terminal Josephson junctions is understood from the motion of the particle trapped in a washboard potential. Furthermore, since qualitatively the same results are obtained even when the absolute value of the order parameter is assumed to be constant, the relative phase of the order parameter, rather than its absolute value, largely contributes to the nonequilibrium dynamics of the dissipative fermionic superfluids.

\section{Conclusions}\label{sec_Conc}
We have investigated the dissipative dynamics of a triad of fermionic superfluids connected via Josephson junctions, following a sudden switch-on of two-body loss. We have formulated the dissipative BCS theory for the Lindblad equation and derived expressions for the Josephson currents. We have found that the superfluid order parameter exhibits a phase rotation, giving rise to three types of dc Josephson currents for $J_{12}, J_{23}$ and $J_{31}$. We have revealed that, when $V_{31}$ is weak, dissipation first induces the \cred{nonequilibrium phase transition} characterized by the emergence of a single finite dc Josephson current, while increasing dissipation leads to the disappearance of the remaining dc Josephson current. We have demonstrated that, when $V_{31}$ is strong, dissipation induces \cred{the nonequilibrium phase transition} characterized by the simultaneous disappearance of all dc Josephson currents.
\cred{The nonequilibrium phase transition} has also been analytically studied by using a simplified model.
Our result can be realized experimentally with ${}^{6}\mathrm{Li}$~\cite{valtolina15,burchianti18,luick20,kwon20,del21,del25} by using the photoassociation techniques~\cite{tomita17,huang25}, since \cred{the nonequilibrium phase transition} has been recently observed in ultracold atoms~\cite{scott19,muniz20}. 
\cred{As we have studied the case in which three superfluids are initially prepared in the BCS ground state with identical interactions, the nonequilibrium dynamics for different initial conditions, such as with different interaction strengths, may give rise to interesting dynamical behavior.}
Though we have focused on the dynamics of the fermionic superfluids, studies for coupled Bose-Einstein condensates are also important. Indeed, since a system in junctions which constitute three Bose-Einstein condensates has been investigated experimentally~\cite{scherer07}, the nonequilibrium dynamics of Bose-Einstein condensates under losses deserves further study.
Moreover, in the loss-only dynamics, it is known that the spectrum of the Liouvillian is obtained solely from that of the corresponding non-Hermitian Hamiltonian~\cite{torres14}. It would be interesting to study how the dynamics is affected by exceptional points, where the spectrum of the non-Hermitian Hamiltonian shows singular behavior in fermionic superfluids~\cite{yamamoto19,shen24,takemori24honey,takemori24asym,takemori25,chin25,takemori26,capecelatro26_super,capecelatro26_andreev}.

\begin{acknowledgments}
  We thank Akihisa Koga and Takashi Mukaiyama for fruitful discussions.
  This work was supported by JSPS
  KAKENHI Grant No.\ JP26KJ1111 (S.T.) and JP25K17327 (K.Y.).
  S.T. was supported by the Sasakawa Scientific Research Grant from the Japan Science Society and JST SPRING, Japan Grant No.\ JPMJSP2180. This work was supported by JSPS Program for Forming Japan's Peak Research Universities (J-PEAKS) Grant No. JPJS00420230008. This work was partly supported by Hirose Foundation, Fujikura Foundation, Toyota Riken Scholar Program, and Support Center for Advanced Telecommunications Technology Research.
\end{acknowledgments}

\section*{Data availability}
The data that support the findings of this article are openly available~\cite{dataset}.

\appendix

\section{\cred{Robustness of the nonequilibrium phase transition for very weak dissipation}} \label{app_weak}

\begin{figure*}[t]
    \centering
    \includegraphics[width=\textwidth]{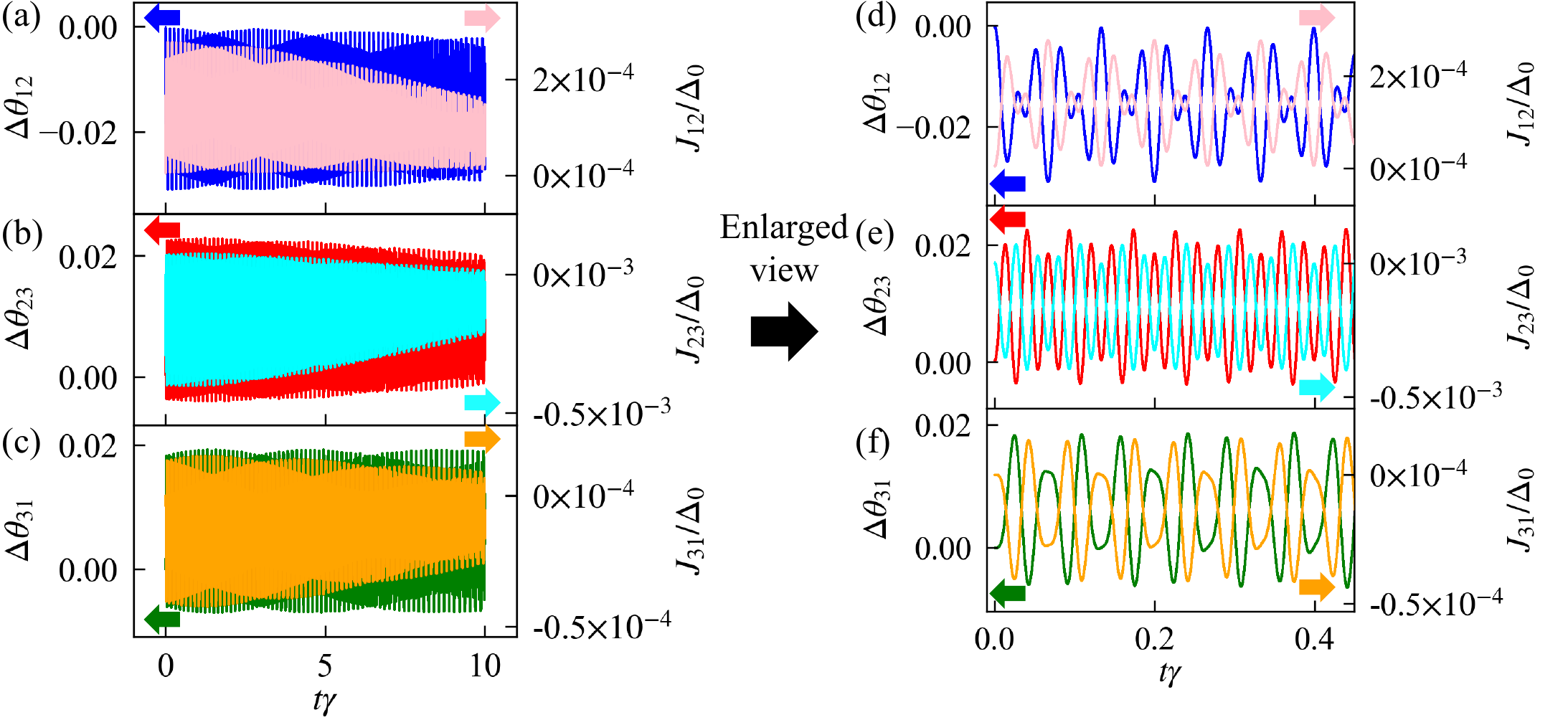}
    \caption{\cred{(a), (b), (c) Phase differences $\Delta\theta_{12}$, $\Delta\theta_{23}$, and $\Delta\theta_{31}$, and corresponding Josephson currents $J_{12}$, $J_{23},$ and $J_{31}$. (d), (e), (f) Enlarged views of the figures.
    The case of very weak loss $\gamma=0.002\Delta_{0}$ is displayed. All Josephson currents have finite dc components. The other parameters are set to the same as those in Fig.~\ref{V310.05g0.8_image}.}}
    \label{fig1_3}
\end{figure*}

\cred{
In this section, we confirm the robustness of the two-step nonequilibrium phase transition for very weak dissipation. In Fig.~\ref{fig1_3}, we perform numerical calculations for Josephson currents $J_{12},J_{23},$ and $J_{31}$ when $\gamma= 0.002\Delta_{0} $ ($\gamma \ll V_{\nu\mu}$).
As shown in Fig.~\ref{fig1_3}, the phase differences $\Delta\theta_{12}$, $\Delta\theta_{23}$, and $\Delta\theta_{31}$ of the order parameters exhibit oscillations with small amplitudes over time, and corresponding Josephson currents have both dc and ac components. 
This behavior is qualitatively the same as that shown in Fig.~\ref{V310.05g0.8_image}. 
We also note that, for very strong dissipation, the particle number rapidly decreases to zero, making it difficult to numerically simulate the transient dynamics in a time window comparable to that studied in Fig.~\ref{V310.05g1.5_image}. 
}

\section{Analytical explanation for the beating behavior under weak dissipation}\label{app_beating}
In this section, we explain, by using the simplified model~\eqref{TriQuench_RCSJ_EOM_eq} for $V_{31}=0$ under the assumption $|\Delta_{1}|\simeq |\Delta_{2}|\simeq |\Delta_{3}|=|\Delta_{1}|_{t=0}$, 
why the beating behavior of the order parameter appears under weak dissipation. Moreover, we show that, if 
the condition $I_{12}=I_{23}$ is satisfied, the order parameter does not exhibit a beating behavior.

\begin{figure}[t]
    \centering
    \includegraphics[width=\linewidth]{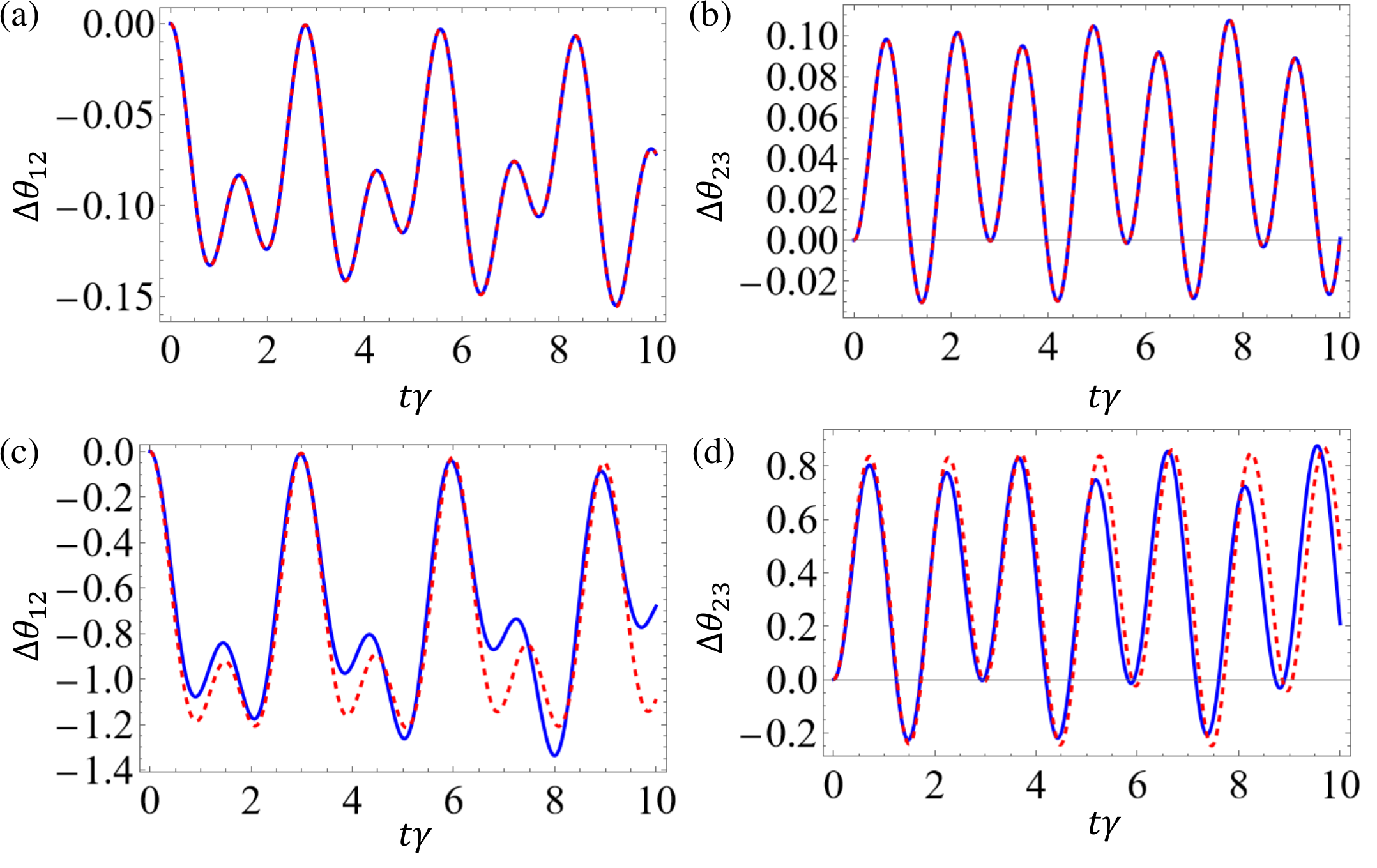}
    \caption{Nonequilibrium dynamics of the phase differences $\Delta\theta_{12}$ and $\Delta\theta_{23}$ following a sudden switch-on of two-body loss. The dissipation strength is set to (a), (b) \cred{$\gamma=0.01\Delta_{0}$} and (c), (d) \cred{$\gamma=0.082\Delta_{0}$}. The blue line shows the solution~\eqref{triquench_app_soleq} while the red line shows the solution obtained by solving the equation of motion~\eqref{TriQuench_RCSJ_EOM_eq} under the assumption $|\Delta_{1}|\simeq |\Delta_{2}| \simeq |\Delta_{3}|=|\Delta_{1}|_{t=0}$. The phase differences exhibit beating behaviors. The other parameters are the same as those in Fig.~\ref{V310_image}.}
    \label{beating_image}
\end{figure}
\begin{figure*}[t]
    \centering
    \includegraphics[width=\textwidth]{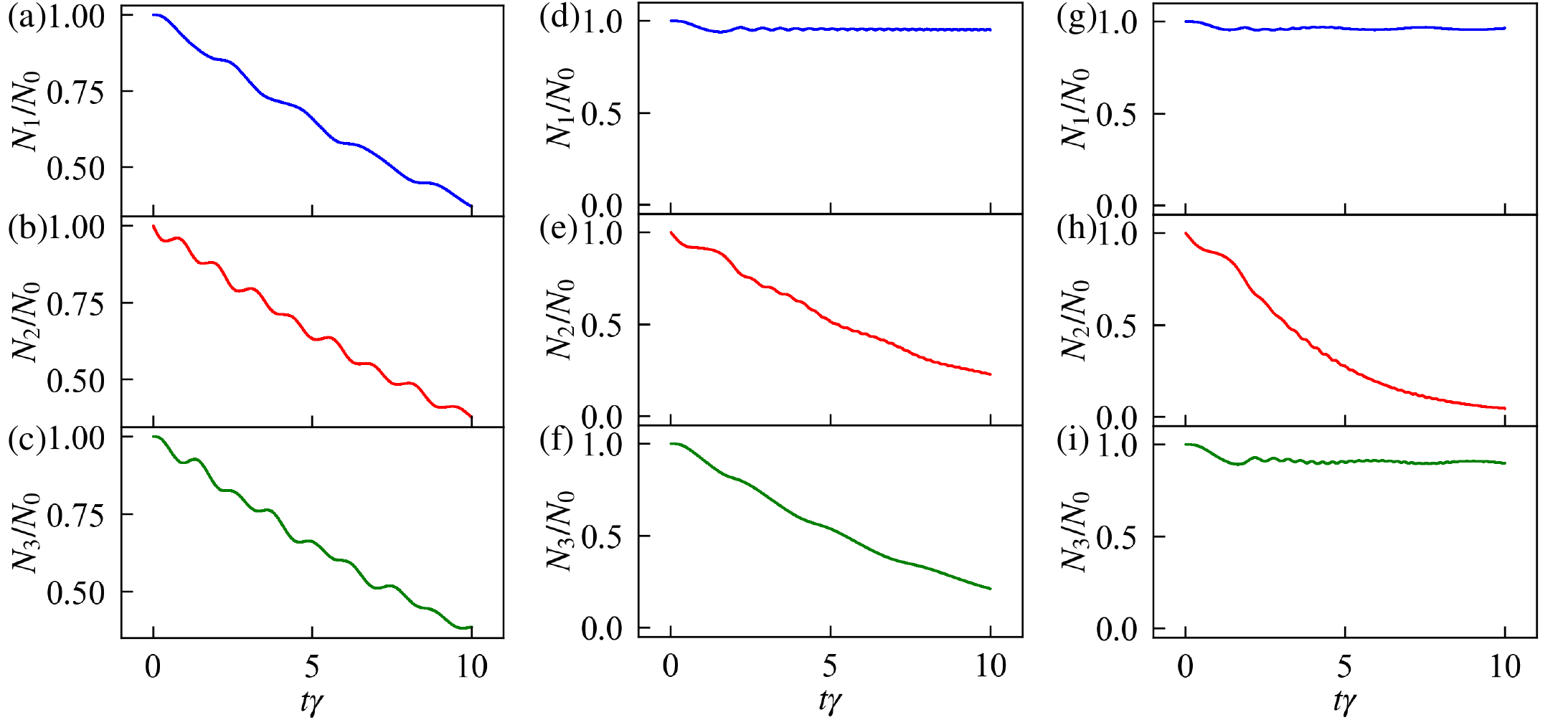}
    \caption{Nonequilibrium dynamics of the particle numbers $N_{1}$, $N_{2}$, and $N_{3}$ following a sudden switch-on of two-body loss. The dissipation strengths are set to (a)-(c) weak, (d)-(f) medium, and (g)-(i) strong values. All the parameters are the same as those in Figs.~\ref{V310.05g0.8_image}-\ref{V310.05g1.5_image}.}
    \label{particlenum_image}
\end{figure*}
We start with the equations of motion for the phase difference, Eqs.~\eqref{TriQuench_phase_a_diff_eq} and \eqref{TriQuench_phase_b_diff_eq}. We assume that the particle is confined around a minimum of the potential~\eqref{TriQuench_RCSJ_pot_eq}, where the minimum is obtained by solving Eqs.~\eqref{triquench_wash_eq1} and ~\eqref{triquench_wash_eq2}. In the $(\Delta\theta_{12},\Delta\theta_{23})$ coordinates, the minimum $(\phi_{12},\phi_{23})$ is given as $\phi_{12}= \arcsin( -C_{0}/3I_{12}),\; \phi_{23}=\arcsin(C_{0}/3I_{23} ).$
Then, Eqs.~\eqref{TriQuench_phase_a_diff_eq} and \eqref{TriQuench_phase_b_diff_eq} are expanded around $(\Delta\theta_{12},\Delta\theta_{23})=(\phi_{12},\phi_{23})$ as
\begin{align}
    \begin{pmatrix}
        \ddot{\Delta\theta_{12}} \\ \ddot{\Delta\theta_{23}}
    \end{pmatrix} &= -WA\begin{pmatrix}
        \Delta\theta_{12} \\ \Delta\theta_{23}
    \end{pmatrix} + W\bm{\Theta},
\end{align}
with
\begin{align}
    &A = \begin{pmatrix}
        2I_{12}\cos\phi_{12} & -I_{23}\cos\phi_{23} \\ -I_{12}\cos\phi_{12} & 2I_{23}\cos\phi_{23}
    \end{pmatrix}, \\
    &\bm{\Theta} = \begin{pmatrix}
        \Theta_{1} \\ \Theta_{2}
    \end{pmatrix}, \\
    &\Theta_{1} = -2I_{12}\sin\phi_{12} +2I_{12}\phi_{12}\cos\phi_{12} \notag\\
    &+ I_{23}\sin\phi_{23} - I_{23}\phi_{23}\cos\phi_{23} -C_{0}, \\
    &\Theta_{2} = I_{12}\sin\phi_{12} -I_{12}\phi_{12}\cos\phi_{12} \notag \\
    &-2I_{23}\sin\phi_{23} + 2I_{23}\phi_{23}\cos\phi_{23} +C_{0},
\end{align}
Since $I_{\nu\mu}\cos\phi_{\nu\mu}>0$, we have $\text{det}(A)= 3I_{12}I_{23}\cos\phi_{12}\cos\phi_{23}>0$, and the equation is rewritten as
\begin{align}
    \begin{pmatrix}
        \ddot{\Delta\theta_{12}'} \\ \ddot{\Delta\theta_{23}'}
    \end{pmatrix} &= -WA\begin{pmatrix}
        \Delta\theta_{12}' \\ \Delta\theta_{23}'
    \end{pmatrix}, \label{TriQuench_app_DEeq} 
\end{align}
where we have introduced
\begin{align}
    \begin{pmatrix}
        \Delta\theta_{12}' \\ \Delta\theta_{23}'
    \end{pmatrix} &=\begin{pmatrix}
        \Delta\theta_{12} \\ \Delta\theta_{23}
    \end{pmatrix} -A^{-1}\bm{\Theta}.
\end{align}
The eigenvalues of matrix $A$ are given as
\begin{equation}
    \lambda_{\pm} = I_{12}\cos\phi_{12} + I_{23}\cos\phi_{23} \pm \sqrt{\Lambda},
\end{equation}
where $\Lambda= (I_{12}\cos\phi_{12})^{2}+(I_{23}\cos\phi_{23})^{2}-I_{12}I_{23}\cos\phi_{12}\cos\phi_{23}$. Using the matrix $P$ that satisfies the relation $A=PDP^{-1}$ with $D=\text{diag}(\lambda_{+},\lambda_{-})$, Eq.~\eqref{TriQuench_app_DEeq} is rewritten as
\begin{equation}
    P^{-1}\begin{pmatrix}
        \ddot{\Delta\theta_{12}'} \\ \ddot{\Delta\theta_{23}'}
    \end{pmatrix} = -DP^{-1}\begin{pmatrix}
        \Delta\theta_{12}' \\ \Delta\theta_{23}'
    \end{pmatrix}.
\end{equation}
Then, we obtain the solution of the differential equation as
\begin{align}
    &P^{-1}\begin{pmatrix}
        \Delta\theta_{12}' \\ \Delta\theta_{23}'
    \end{pmatrix} =\begin{pmatrix}
        \cos (\sqrt{W\lambda_{+}}t) & 0 \\ 0 & \cos(\sqrt{W\lambda_{-}}t)
    \end{pmatrix}\begin{pmatrix}
        a_{1} \\ b_{1} 
    \end{pmatrix} \notag \\
    &+ \begin{pmatrix}
        \sin (\sqrt{W\lambda_{+}}t) & 0 \\ 0 & \sin(\sqrt{W\lambda_{-}}t)
    \end{pmatrix}\begin{pmatrix}
        a_{2} \\ b_{2} 
    \end{pmatrix},
\end{align}
where $a_{i}$ and $b_{i}\; (i=1,2)$ are the constant of integration. Using the initial conditions $(\Delta\theta_{12},\Delta\theta_{23})=(0,0)$ and $(\Delta\dot{\theta}_{12},\Delta\dot{\theta}_{23})=(0,0)$,
we obtain
\begin{align}
    \begin{pmatrix}
        \Delta\theta_{12} \\ \Delta\theta_{23}
    \end{pmatrix} &= A^{-1}\bm{\Theta} \notag\\ \displaybreak[2]
    &-P\begin{pmatrix}
        \cos (\sqrt{W\lambda_{+}}t) & 0 \\ 0 & \cos(\sqrt{W\lambda_{-}}t)
    \end{pmatrix}P^{-1}A^{-1}\bm{\Theta}.  \label{triquench_app_soleq}
\end{align}
Since we have $\lambda_{+}\neq \lambda_{-}$ if $I_{12}\neq I_{23}$, Eq.~\eqref{triquench_app_soleq} demonstrates that the phase differences $\Delta\theta_{12}$ and $\Delta\theta_{23}$ exhibit beating behavior. When $I_{12}=I_{23}$, the solution~\eqref{triquench_app_soleq} reduces to a single-frequency oscillation.

In Fig.~\ref{beating_image}, we show the phase differences following Eq.~\eqref{triquench_app_soleq}. For sufficiently weak dissipation (\cred{$\gamma=0.01\Delta_{0}$}), the solution~\eqref{triquench_app_soleq} agrees well with the numerical solution obtained by solving the equation of motion~\eqref{TriQuench_RCSJ_EOM_eq} [see Fig.~\ref{beating_image} (a) and (b)]. In contrast, for slightly larger (but still weak) dissipation (\cred{$\gamma=0.082\Delta_{0}$}), the solution~\eqref{triquench_app_soleq} deviates from the numerical results [see Fig.~\ref{beating_image} (c) and (d)], because the particle explores regions slightly away from the minimum of the potential as dissipation increases. In both cases (\cred{$\gamma=0.01\Delta_{0}$ and $0.082\Delta_{0}$}), the order parameter exhibits a beating behavior.

\section{Dynamics of the particle number}\label{app_particle}
In this section, we show the dynamics of the particle number obtained by numerically solving Eq.~\eqref{tri_bloch_eq}. We consider the cases where the two-body loss is weak, medium, and strong. The parameters are set to the same values as in Figs.~\ref{V310.05g0.8_image}-\ref{V310.05g1.5_image}. For weak dissipation shown in Figs.~\ref{particlenum_image}(a)-(c), all particle numbers $N_{1},N_{2},$ and $N_{3}$ decrease with oscillations. For medium dissipation shown in Figs.~\ref{particlenum_image}(d)-(f), the decrease of particle number $N_{1}$ is slower than that of $N_{2}$ and $N_{3}$, which is consistent with the fact that the dc Josephson currents $J_{12}^{\text{dc}}$ and $J_{31}^{\text{dc}}$ vanish (see Fig.~\ref{V310.05g1.11_image}). Finally, for strong dissipation, the particle number $N_{2}$ decreases faster than that of system 1 and system 3 [see Figs.~\ref{particlenum_image}(g)-(i)]. This is consistent with the fact that all dc Josephson currents vanish.

\section{\cred{Robustness of the nonequilibrium phase transition for different values of $U_{R}$ and $V_{\nu\mu}$}} \label{app_UV}

\begin{figure*}[tb]
    \centering
    \includegraphics[width=\textwidth]{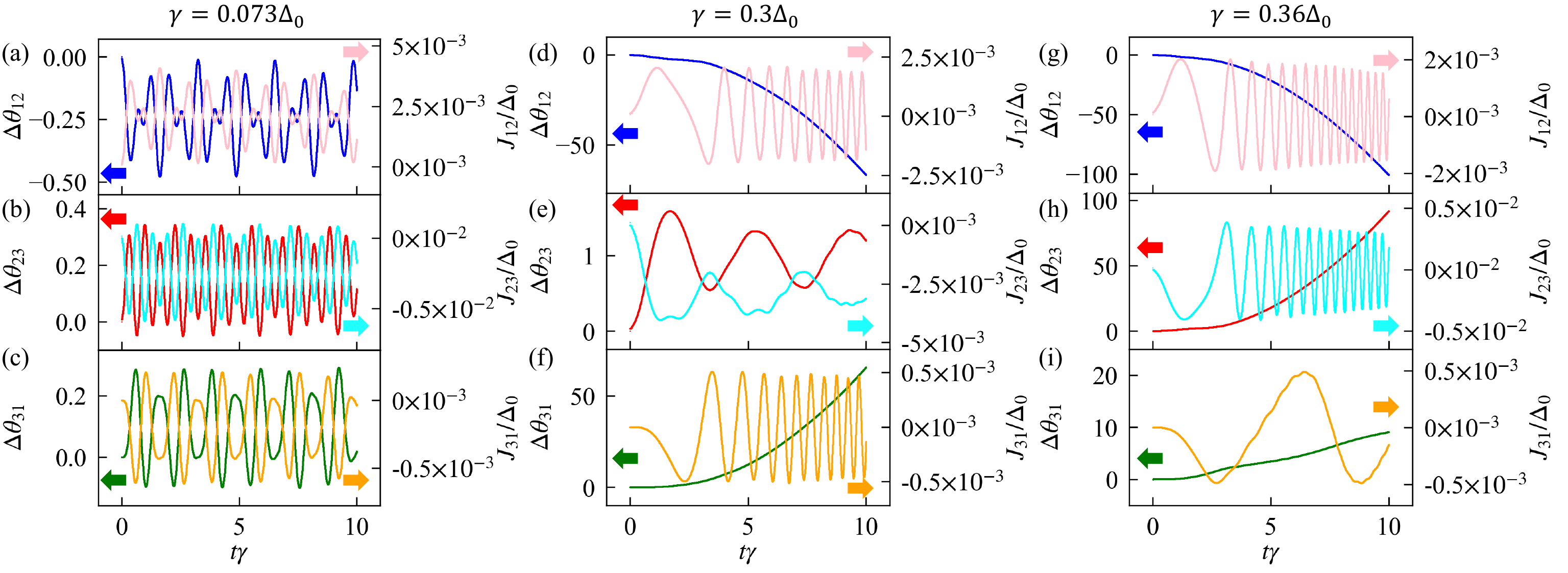}
    \caption{\cred{Phase differences $\Delta\theta_{12}$, $\Delta\theta_{23}$, and $\Delta\theta_{31}$, and corresponding Josephson currents $J_{12}$, $J_{23},$ and $J_{31}$ for $(U_{R},V_{12}) = (4.8\Delta_{0},0.048\Delta_{0})$ are plotted. 
    The dissipation strength is set to (a)-(c) $\gamma=0.073\Delta_{0}$, (d)-(f) $\gamma=0.3\Delta_{0}$, and (g)-(i) $\gamma=0.36\Delta_{0}$. For weak dissipation [(a)-(c)], all Josephson currents have finite dc components. For medium dissipation [(d)-(f)], only $J_{23}$ has a finite dc component. For strong dissipation [(g)-(i)], all Josephson currents do not have finite dc components. The other parameters are set to the same as those in Figs.~\ref{V310.05g0.8_image}-\ref{V310.05g1.5_image}.}}
    \label{fig1_4_abc}
\end{figure*}

\cred{
In this section, we confirm the robustness of the two-step nonequilibrium phase transition across different parameter regimes. In Figs.~\ref{V310.05g0.8_image}-\ref{V310.05g1.5_image}, we have performed numerical calculations for $(U_{R}, V_{12}) = (3.1\Delta_{0}, 0.02\Delta_{0})$, demonstrating that the two-step nonequilibrium phase transition occurs for weak $V_{31}$.
We further examine the robustness of this result by performing additional calculations for four parameter sets: $(U_{R}, V_{12}) = (4.8\Delta_{0}, 0.048 \Delta_{0}),\;(2.8\Delta_{0}, 0.015\Delta_{0}),\;(4.8\Delta_{0}, 0.015\Delta_{0}),\;$ and $(2.8\Delta_{0}, 0.022\Delta_{0})$. All other parameters are kept the same as those used in Figs.~\ref{V310.05g0.8_image}-~\ref{V310.05g1.5_image}.
}

\cred{
As an illustrative example, we consider the case with an interaction strength and a tunneling amplitude $(U_{R}, V_{12}) = (4.8\Delta_{0},0.048\Delta_{0})$. In Figs.~\ref{fig1_4_abc} (a)-(c), we show the phase difference $\Delta\theta_{\nu\mu}$ of the order parameters and the corresponding Josephson current for weak dissipation ($\gamma = 0.073\Delta_{0}$). The phase differences show an oscillation around nonzero constant values, and the Josephson currents have both dc and ac components. For medium dissipation ($\gamma = 0.3\Delta_{0}$) as shown in Figs.~\ref{fig1_4_abc} (d)-(f), the phase difference $\Delta\theta_{23}$ oscillates with small amplitudes over time, whereas $|\Delta\theta_{12}|$ and $|\Delta\theta_{31}|$ rapidly increase over time. Correspondingly, the Josephson current $J_{23}$ retains both ac and dc components, while the dc components of $J_{12}$ and $J_{31}$ vanish. Moreover, for strong dissipation ($\gamma= 0.36\Delta_{0}$), all the absolute values of the relative phase $|\Delta\theta_{\nu\mu}|$ rapidly increase and all dc components of $J_{\nu\mu}$ vanish [see Figs.~\ref{fig1_4_abc} (g)-(i)]. These results signal the nonequilibrium phase transition and are qualitatively the same as Figs.~\ref{V310.05g0.8_image}-\ref{V310.05g1.5_image}.
We also confirm that the other parameters give qualitatively the same behavior.
These results highlight that varying the interaction strength $U_R$ and the tunneling amplitude $V_{\nu\mu}$ does not qualitatively affect the nonequilibrium phase transition of the dissipative fermionic superfluids.
}


%

\end{document}